\documentclass[twocolumn,fleqn, resetfootnote]{aastex701}

\usepackage{subcaption} 
\usepackage{enumitem} 

\newcommand{\se}[1]{Section \ref{sec:#1}}

\newcommand{\Fig}[1]{Fig.~\ref{fig:#1}}
\newcommand{\Figs}[1]{Figs.~\ref{fig:#1}}

\newcommand{\be}{\begin{equation}}
\newcommand{\ee}{\end{equation}}
\newcommand{\bad}{\begin{equation} \begin{aligned}}
\newcommand{\ead}{\end{aligned} \end{equation}}

\newcommand{\Msun}{M_\odot}

\newcommand{\g}{\,{\rm g}}

\newcommand{\Mv}{M_{\rm vir}}
\newcommand{\Ms}{M_{\star}}

\newcommand{\Mb}{M_{\rm b}}

\newcommand{\Mdm}{M_{\rm dm}}

\newcommand{\ms}{m_\star}

\newcommand{\rv}{r_{\rm vir}}
\newcommand{\Reff}{R_{\rm e}}
\newcommand{\Rein}{R_{\rm e,in}}
\newcommand{\rhalf}{r_{\rm 1/2}}

\newcommand{\abar}{a_{\rm bar}}

\newcommand{\vrot}{v_{\rm rot}}

\newcommand{\Sigmag}{\Sigma_{\rm g}}
\newcommand{\Sigmas}{\Sigma_\star}

\usepackage{subcaption}
\usepackage{amsmath}
\usepackage{array}
\usepackage{natbib}
\usepackage{multirow}
\begin{document}

\title{Ubiquitous nuclear disks and bars embedded in massive galaxies in the cosmic morning}

\author[orcid=0009-0000-0943-2107]{Jianyuan Luo}
\affiliation{Department of Astronomy, School of Physics, Peking University, Beijing 100871, China}
\email{2401110279@stu.pku.edu.cn}

\author[orcid=0000-0001-6115-0633,sname='Jiang']{Fangzhou Jiang}
\altaffiliation{Corresponding author}
\affiliation{Kavli Institute for Astronomy and Astrophysics, Peking University, Beijing 100871, China}
\affiliation{Department of Astronomy, School of Physics, Peking University, Beijing 100871, China}
\email[show]{\href{mailto:fangzhou.jiang@pku.edu.cn}{fangzhou.jiang@pku.edu.cn}}

\author[orcid=0000-0001-8405-2921]{Jinning Liang}
\affiliation{Department of Astronomy, School of Physics, Peking University, Beijing 100871, China}
\email{}

\author[orcid=0000-0001-9215-7053]{Boris S. Kalita}
\affiliation{Kavli Institute for Astronomy and Astrophysics, Peking University, Beijing 100871, China}
\email{}

\author[orcid=0000-0002-4328-538X]{Pinsong Zhao}
\affiliation{Kavli Institute for Astronomy and Astrophysics, Peking University, Beijing 100871, China}
\email{}

\author[orcid=0000-0002-4569-9009]{Jinyi Shangguan}
\affiliation{Kavli Institute for Astronomy and Astrophysics, Peking University, Beijing 100871, China}
\affiliation{Department of Astronomy, School of Physics, Peking University, Beijing 100871, China}
\email{}

\author[orcid=0009-0003-4862-2925]{Bingcheng Jin}
\affiliation{Department of Astronomy, University of Michigan, Ann Arbor, MI 48109, USA}
\email{}

\author[orcid=0000-0002-0182-1973]{Zeyu Gao}
\affiliation{Kavli Institute for the Physics and Mathematics of the Universe (WPI), The University of Tokyo, Kashiwa, Japan}
\affiliation{Department of Astronomy, School of Physics, Peking University, Beijing 100871, China}
\email{}

\author[orcid=0000-0002-9593-8274]{Weichen Wang}
\affiliation{Department of Physics, Universita degli Studi di Milano-Bicocca, Piazza della Scienza, 3, Milano,
I-20126, Italy}
\email{}

\author[orcid=0000-0001-6947-5846]{Luis C. Ho}
\affiliation{Kavli Institute for Astronomy and Astrophysics, Peking University, Beijing 100871, China}
\affiliation{Department of Astronomy, School of Physics, Peking University, Beijing 100871, China}
\email{}

\author[orcid=0000-0003-0939-9671]{Yingjie Peng}
\affiliation{Kavli Institute for Astronomy and Astrophysics, Peking University, Beijing 100871, China}
\affiliation{Department of Astronomy, School of Physics, Peking University, Beijing 100871, China}
\email{}

\begin{abstract}
The morphology of the central regions of high-redshift galaxies remains relatively unexplored, while recent case studies suggest that giant disks can emerge without first developing a prominent central spheroid. Here, we investigate the inner structures of a mass-complete sample of 45 massive galaxies at $z=3$ in the TNG50 simulation. Through double-Sérsic profile decomposition, isodensity ellipse fitting, intrinsic three-dimensional shape measurements, and stellar kinematics, we find that these central structures are ubiquitously flattened and rotation-supported with low Sérsic indices ($n \lesssim 1$). This indicates that the central regions are predominantly nuclear disks and bars rather than bulges. Tracking their evolution, we show that these nuclear disks and bars emerge during gas-rich compaction and do not transform into spheroidal and dispersion-dominated systems until $z \lesssim 1$, before which extended stellar disks have already developed. Our results suggest that massive disks can assemble around rotation-supported centers and bulge-deficient massive galaxies may be a common outcome of galaxy evolution.
\end{abstract}



\section{Introduction}
\label{sec:intro}

Galaxy morphology encodes valuable information about galaxy formation and evolution \citep[e.g.,][]{Kormendy04,Conselice14}. 
The James Webb Space Telescope (JWST) has enabled detailed studies of galaxy morphology out to the epoch of cosmic reionization \citep[e.g.,][]{Kartaltepe23, Huertas-Company24, Lee24, Yu26}. 
At the same time, modern cosmological simulations have become capable of reproducing galaxies spanning a wide range of morphological types and have reached resolutions that, in principle, allow predictions of fine morphological structures across cosmic time \citep[e.g.,][]{Tacchella19, ForouharMoreno26}.
A major frontier in our understanding of galaxy formation is therefore determining whether cosmological simulations reproduce the correct galaxy morphologies at the appropriate epochs. 
Most morphological studies, particularly those comparing simulations with observations, have focused on global properties such as overall shape or the relative dominance of bulge and disk components \citep[e.g.,][]{Costantin23, Pandya24, Vega-Ferrero24, Gong25}, while paying comparatively little attention to the detailed internal structures of galaxies.
In practice, an accurate description of galaxy morphology often requires multiple structural components, each of which may encode information about distinct evolutionary processes or stages \citep[e.g.,][]{Huang13, Salo15, Bland-Hawthorn16}.
Such structural complexity suggests that simple global morphological classifications are insufficient and that a more detailed examination of galactic substructures is necessary.
The central regions of galaxies are of particular interest. The galactic centers exhibit remarkable morphological diversity, including bulges, nuclear disks, bars, star clusters, and spiral structures \citep[e.g.,][]{Boker02, Erwin24, Stacey25, LeConte26}. More importantly, central structures are closely related to key physical processes such as gas inflow, central star formation, quenching, and AGN activity \citep[e.g.,][]{Schultheis25, Garland26}. Understanding the nature of central structures is therefore essential for establishing the link between galaxy morphology and the physical mechanisms driving galaxy evolution.

JWST has revealed a growing population of high-redshift galaxies with well-developed disk morphologies \citep[e.g.,][]{Robertson23, Sun24, Huertas-Company25, Yu25}. Cosmological simulations have revealed a popular theoretical picture for disk formation, known as the ``gas-rich compaction" \citep[e.g.,][]{Dekel14, Ceverino15}. In this picture, galaxies experience a phase of rapid gas inflow and central gas accumulation, leading to enhanced central star formation and the formation of a compact stellar component. This compact component is often interpreted as the progenitor of a bulge, around which galaxies may subsequently regrow extended disks during post-compaction evolution.
However, a recent study revealed a giant disk at $z\approx3$, referred to as the ``Big Wheel'' \citep{Wang25}, whose central bulge-like component contributes at most a few percent of the stellar mass \citep{Quadri26}, comparable to the bulge fractions observed in nearby ``bulgeless giant galaxies'' \citep{Kormendy10}. 
In a theoretical follow-up study, \citet{Jiang25} investigated the formation of giant bulge-deficient galaxies from the perspective of the galaxy-dark-matter-halo connection and their large-scale environments, using simulated analogs of the Big Wheel in cosmological simulations.
While the focus of that work was to identify the halo and environmental conditions that promote the formation of such systems, a serendipitous finding was that the central components of all simulated giant disks exhibit low S\'ersic indices, $n\la1$, indicative of disks or bar-like structures rather than bulges. 
Intriguingly, the central component of the observed Big Wheel galaxy also has a similarly low S\'ersic index of $n=0.72$ \citep{Jiang25}. Another statistical study based on the COSMOS-Web survey performed double-S\'ersic decompositions with fixed 
inner and outer S\'ersic indices of 4 and 1, respectively, on galaxies ranging from $0<z<10$. Under these constraints, the analysis also identified a population of massive, bulge-deficient galaxies with low bulge-to-total ratios at high redshift \citep{Yang26}.  
These studies motivate us to investigate a complete sample of simulated massive galaxies at high redshift and systematically characterize the morphology of their central structures.

The remainder of this Letter is organized as follows. 
In \se{method}, we describe the high-redshift massive galaxy sample from cosmological simulations, and the methods used to characterize their inner structures. 
In \se{StatisticsandMock}, we present the statistics of these structures at $z=3$, showing that most galaxies host distinct central components that are disky or bar-like. 
In \se{evolution}, we investigate the origin and evolution of these structures in the context of gas-rich compaction and post-compaction morphological evolution. 
Finally, we summarize our conclusions in \se{conclusion}.

\begin{figure*}[t]
  \centering
  \includegraphics[width=0.95\linewidth]{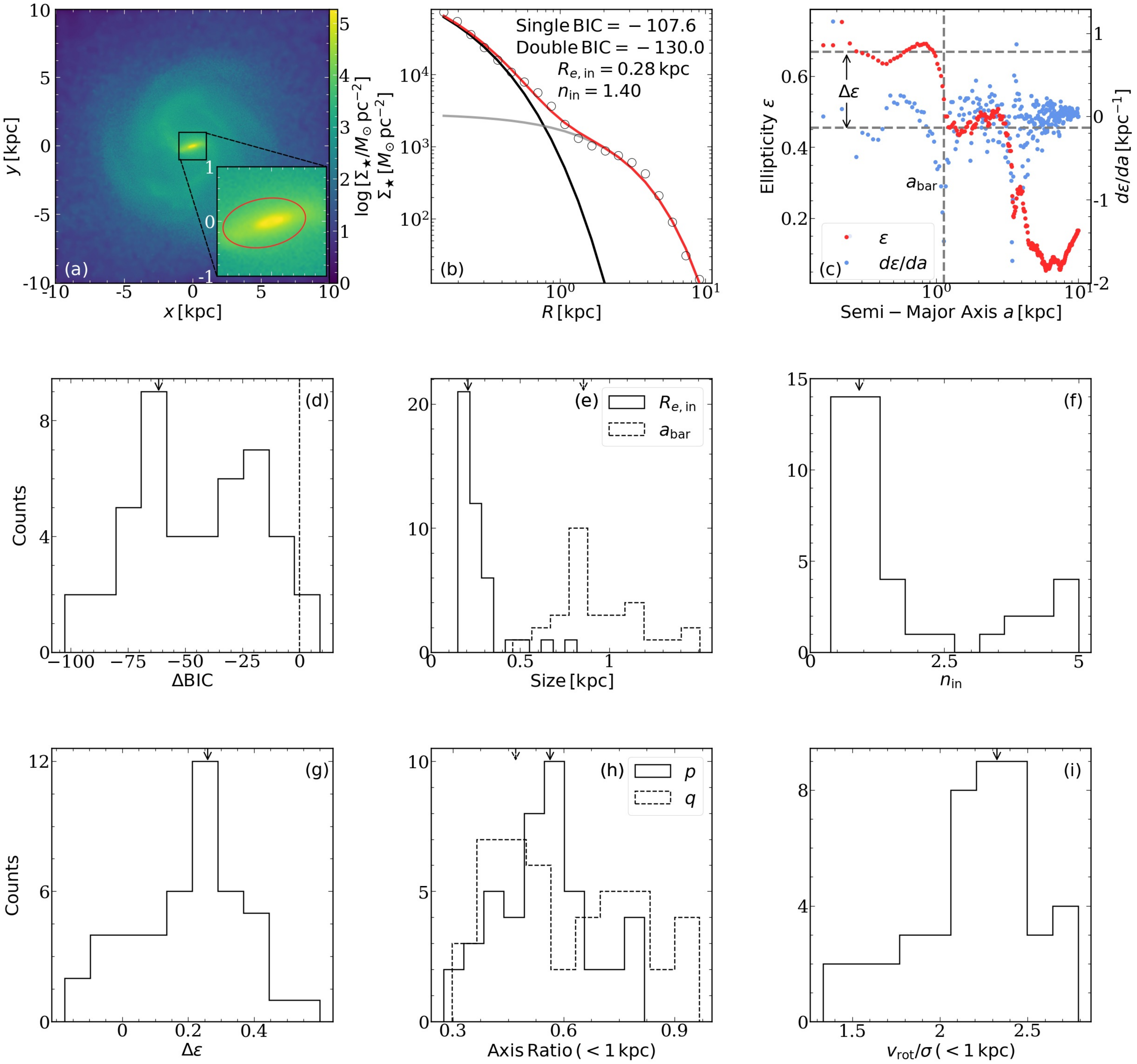} 
  \caption{
  {\bf Case study (top row) and statistical properties (middle and bottom rows) of the compact central structures in the complete sample of 45 massive ($\Ms>10^{10.5}\Msun$) TNG50 galaxies at $z=3$.}
  (a) Face-on stellar surface density map of galaxy ID 34088. 
  The lower right panel shows a zoom-in of the central region, revealing a compact elongated structure. The red ellipse indicates the initial guess for the isodensity ellipse fitting performed with the \texttt{photutils.isophote} module in \texttt{Astropy}. 
  (b) Double-Sérsic fit to the stellar surface density profile. 
  Circles show the simulation measurements, while the black, grey, and red curves represent the best-fit inner component, outer component, and total model, respectively. 
  The double-Sérsic model is strongly favored over a single-Sérsic model according to the Bayesian Information Criteria (BIC).
  (c) Ellipticity profile as a function of semi-major axis. The red points show the ellipticity $\epsilon$ and the blue points show its first derivatives with respect to the semi-major axis $\rm d\epsilon/\rm da$.
  The bar radius, $\abar$, is defined as the location of the minimum of $\rm d\epsilon/\rm da$ (vertical dashed line), while the ellipticity drop, $\Delta\epsilon$ (horizontal dashed lines), quantifies the strength of the compact elongated structure.
  \quad
  Panels (d)–(i) summarize the morphological properties of the complete galaxy sample: 
  (d) The difference in BIC between the single- and double-Sérsic fits, with negative values favoring a distinct inner component, where the vertical dashed line marks $\Delta{\rm BIC}=0$; 
  (e) effective radius of the inner Sérsic component, $R_{\rm e,in}$, and bar radius, $\abar$; 
  (f) inner Sérsic index, $n_{\rm in}$; 
  (g) ellipticity drop, $\Delta\epsilon$; 
  (h) intrinsic axis ratios, $q=b/a$ and $p=c/b$; and 
  (i) rotation-to-dispersion ratio, $\vrot/\sigma$. Quantites in panels (g)-(i) are measured within the central 1 kpc.
  }
  \label{fig:ExampleandStatistics}
\end{figure*}

\section{Method} 
\label{sec:method}

In this section, we describe the simulation sample and the methods used to identify and characterize compact inner components in massive high-redshift galaxies. We begin by demonstrating that such galaxies ubiquitously host distinct inner components and then proceed to quantify their morphological and kinematic properties through two-dimensional and three-dimensional analyses.

\subsection{Simulation sample selection} 
\label{sec:sample}

We use the TNG50 simulation from the IllustrisTNG project \citep{Pillepich19}, which follows galaxy formation in a periodic volume of $51.7\,{\rm cMpc}^3$ with a baryonic mass resolution of $8.5\times10^4\,M_\odot$. The gravitational softening length of collisionless particles is 575 comoving pc at $z>1$ and 288 physical pc thereafter. Gas cells have a minimum softening length of 72 comoving pc, while star-forming gas has a typical cell size of 140 comoving pc. This combination of volume and resolution makes TNG50 well suited for studying galaxy structure while retaining a statistically meaningful sample.

JWST observations have revealed a significant population of massive disky galaxies at $z\ga 3$ \citep[e.g.,][]{Ferreira23,Nelson23}, including systems that are well-developed disks with spiral structures and bars \citep[e.g.,][]{Wang25,Umehata24}. 
Motivated by these discoveries, we select the complete sample of 45 TNG50 central galaxies with stellar masses $\Ms>10^{10.5}\Msun$ at $z=3$ and trace their evolution across cosmic time. 
We do not impose an additional morphological selection for disk galaxies, as most of these galaxies already exhibit significant disk components. 

For face-on and edge-on visualizations, we adopt the subhalo position from the public {\tt SubFind} catalog\footnote{\url{https://www.tng-project.org/data/docs/specifications/\#sec2b}} as the galaxy center and compute the stellar angular-momentum vector using all star particles within 5 times half-stellar-mass radius $r_{1/2}$ -- particle coordinates are then transformed into a frame whose $z$-axis is aligned with this angular-momentum vector.

\subsection{Identifying compact inner components} 
\label{sec:SersicFitting}

We characterize the face-on stellar surface-density distributions of the simulated galaxies using S\'ersic profiles \citep[][]{Sersic63,Ciotti99}, 
\be\label{eq:Sersic}
\Sigma(R)=\Sigma_{\rm e} e^{-b_{\rm n} \left[ \left( R/\Reff \right)^{1/n} - 1 \right]}
\ee
where $\Reff$ is the effective radius, $n$ is the S\'ersic index, $b_{\rm n}=2n-1/3+4/405n$, and $\Sigma_{\rm e}$ is the surface density at $\Reff$. 
Stellar surface-density profiles of simulated galaxies, $\Sigma_{\rm sim}(R_i)$, are measured in 20 logarithmically spaced radial bins, with $R_i$ extending from the collisionless softening length to $\max\{5\rhalf,0.1\rv\}$, where $\rv$ is the virial radius of host halo. 

We fit each profile with both a single-S\'ersic model and a double-S\'ersic model using \texttt{lmfit}\footnote{\url{https://lmfit.github.io/lmfit-py/intro.html}}.
We assess the need for two components using the Bayesian Information Criterion (BIC),
$\text{BIC} = N\ln(\chi^2/N) + N_{\text{var}}\ln(N)$,
where $N$ is the number of data points, $N_{\text{var}}$ is the number of free parameters, and $\chi^2 = \sum_{i=1}^{N}\{[\log\Sigma_\text{sim}(R_i)-\log\Sigma(R_i)]/\log\Sigma(R_i)\}^2$.
A galaxy is considered to host a distinct inner component if the double-Sérsic fit yields a lower BIC than the single-Sérsic fit.
We consider three free parameters for each S\'ersic component, $\log\Sigma_{\rm e}$, $\Reff$, and $n$, and allow the S\'ersic index to vary from 0.3 to 5.  
We determine the best-fit parameters by minimizing $\chi^2$.
For double-S\'ersic fits, the component with the smaller $\Reff$ in the double-S\'ersic fit is identified as the inner component. 

\subsection{Bar identification from isodensity fitting} \label{sec:EllipseFitting}

In addtion to 1D profiles, we also analyze 2D stellar surface density maps in face-on views.
Many galaxies exhibit elongated central structures reminiscent of bars. 
To quantify their morphology, we perform isodensity ellipse fitting, a standard technique for bar identification and characterization \citep[e.g.,][]{Aguerri09, Kim21}. 
Ellipses are fitted to isodensity contours of the logarithmic surface density maps using the  \texttt{photutils.isophote} module in \texttt{Astropy}\footnote{\url{https://photutils.readthedocs.io/en/latest/user_guide/isophote.html}}. 
The initial guess of the inner most ellipse, as required input of the module, is defined by visual inspection around the bright central region, as illustrated in the upper row \Fig{ExampleandStatistics}. 

Unlike lower-redshift bars, these compact high-redshift structures do not typically show an ellipticity profile that rises to a plateau before declining (e.g., \citet{LeConte24}, Fig.1; \citet{Lyu26}, Fig.8). 
Instead, the ellipticity $\epsilon$ remains roughly constant in the center and then drops with increasing radius.
We define the bar radius, $a_{\text{bar}}$, as the location of the steepest ellipticity decline, corresponding to the minimum of $\mathrm{d}\epsilon/\mathrm{d}a$ along the semi-major axis \citep{Athanassoula02}. 
An ellipticity drop $\Delta\epsilon$ is defined as the difference between the ellipticity in the central flat region within $a_{\text{bar}}$ and the value at the end of the decline.
The ellipticity drop across this transition provides a metric of bar strength.

\subsection{Three-dimensional shape} 
\label{sec:3dshape}

\begin{figure*}[t]
  \centering
  \includegraphics[width=0.87\linewidth]{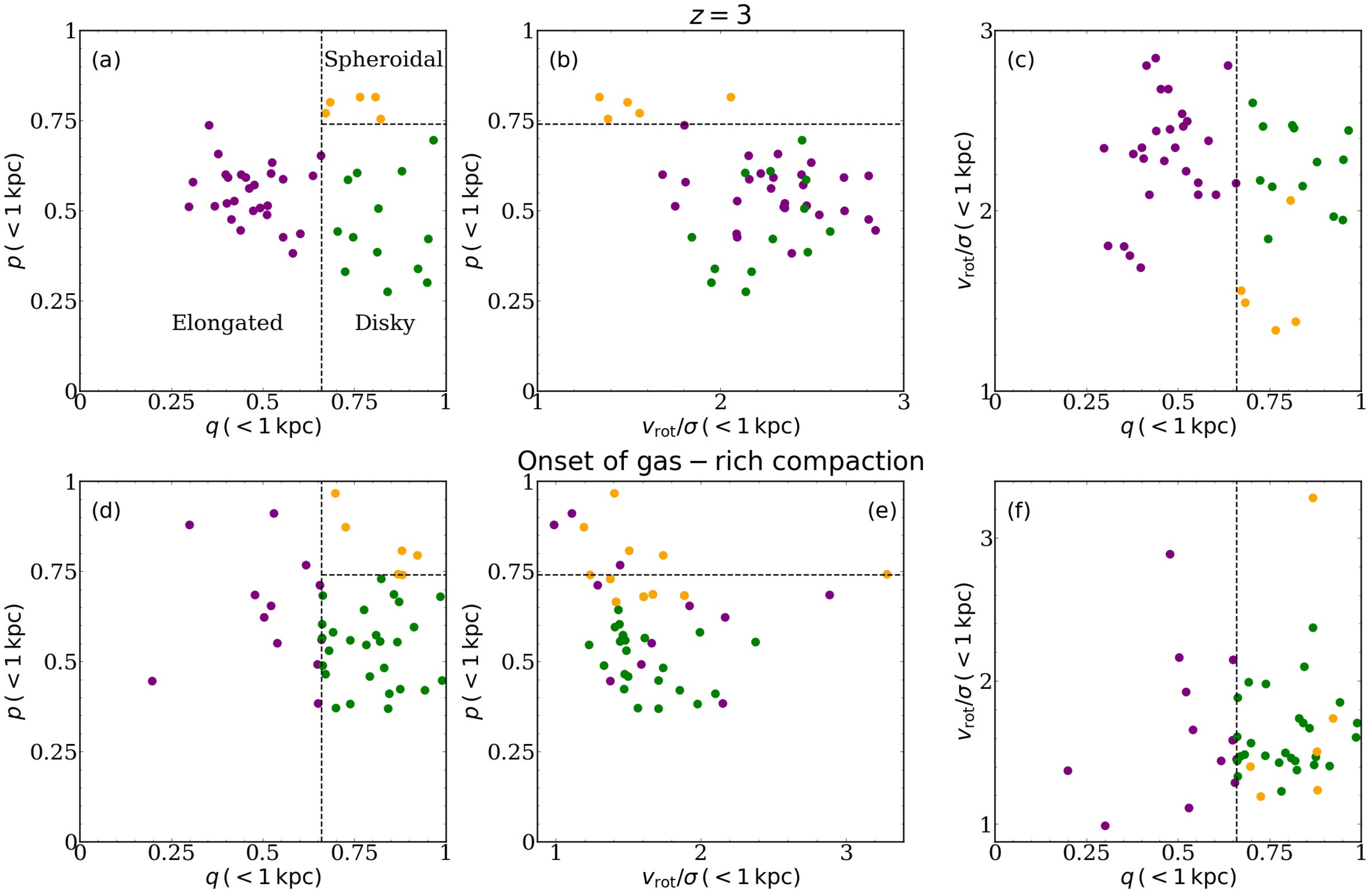} 
  \caption{
{\bf Morphological classification at $z=3$ (top row) and at the onset of gas-rich compaction (bottom row) in the three-dimensional parameter space of the intrinsic axis ratios $q=b/a$, $p=c/b$, and the rotation-to-dispersion ratio $\vrot/\sigma$, all measured within the central 1 kpc.}
The onset of compaction is defined as the epoch when the central 1 kpc first becomes baryon dominated (see \se{evolution} and \Fig{compaction}). 
Galaxies are grouped into three clusters—spheroidal, disky, and elongated—using the \texttt{KMeans} algorithm. 
Dashed lines indicate the approximate cluster boundaries at $q\approx0.7$, $p\approx0.7$, and $\vrot/\sigma\approx1.6$. 
The central regions of many systems are already flattened and rotation supported at the onset of compaction, and evolve toward smaller axis ratios and larger $\vrot/\sigma$ as compaction proceeds.
}
  \label{fig:classification}
\end{figure*}

Not all galaxies exhibit obvious bar-like features, and 2D projections cannot fully capture the intrinsic 3D structure of the inner component. 
We therefore measure its 3D shape using the iterative shape-tensor measurement of \citep{Tomassetti16}. 
Starting from an initial spherical volume of radius $r$, we compute the shape tensor using all the star particles (of mass $\ms$) within it,
\be\label{eq:shape}
S_{ij} = \sum \ms x_i x_j/\sum \ms,
\ee
where $x_i$ and $x_j$ are Cartesian components of particle position vectors relative to the galaxy center, determine the eigenvalues ($a\ge b\ge c$) and eigenvectors, rescale the ellipsoid to set the major axis to $a=r$, rotate particle coordinates into the principal-axis frame, exclude particles outside the ellipsoid, and iterate until the axis ratios $q=b/a$ and $p=c/b$ converge to within 0.01.

\subsection{Kinematics} 
\label{sec:kinematics}

We further characterize the inner morphology kinematically by measuring the ratio of rotational velocity to velocity dispersion, $v_{\rm rot}/\sigma$, within the same radius $r$ adopted for the shape-tensor analysis (\se{3dshape}). 
We estimate $\sigma$ by first computing the in-plane velocity dispersion for each star particle from the velocities of its 16 nearest neighbors, and then taking the mass-weighted average over all particles within $r$.
We compute $v_{\rm rot}$ as the mass-weighted average in-plane tangential velocities over all star particles within $r$.

\subsection{Morphological classification} 
\label{sec:classification}
We classify the galaxies into different morphologies based on the three-dimensional parameter space spanned by the axis ratios $q$, $p$ (\se{3dshape}) and the rotation-to-dispersion ratio $v_{\rm rot}/\sigma$ (\se{kinematics}). 
We adopt the \texttt{KMeans} algorithm in \texttt{scikit-learn}\footnote{\url{https://scikit-learn.org/stable/modules/generated/sklearn.cluster.KMeans.html}}, which divides the 45 samples into 3 clusters by minimizing the distance between data points and the cluster centers. The algorithm is run 10 times with different random initializations and the solution with the best minimization is selected.
The cluster with large $q$ and $p$ alongside small $v_{\rm rot}/\sigma$ values is referred to as spheroidal. The cluster with large $q$, smaller $p$, and large $v_{\rm rot}/\sigma$ is identified as disky morphology. The cluster with small $q$ and large $v_{\rm rot}/\sigma$ is considered elongated.
We will present the classification results in \Fig{classification}.

\section{Ubiquitous nuclear disks and bars in the cosmic morning} 
\label{sec:StatisticsandMock}
A comprehensive statistical characterization of the central regions of the massive galaxies at $z=3$ is presented in \Figs{ExampleandStatistics}–\ref{fig:classification}. A distinct compact inner component is ubiquitous: 43 of the 45 galaxies are better described by a double-Sérsic model than by a single-Sérsic model according to the BIC. 
The inner components have a median Sérsic index of $n_{\rm in}\approx1$, with 72 per cent having $n_{\rm in}<1.5$. 
Their effective radii, $\Rein$, are typically sub-kiloparsec, ranging from 0.2 to 0.8 kpc, with the vast majority having $\Rein<0.5$ kpc. 

About 66 per cent of the sample (30/45) exhibits a pronounced ellipticity drop of $\Delta\epsilon>0.15$, a characteristic signature of a bar-like structure\footnote{The threshold $\Delta\epsilon=0.15$ is chosen based on visual inspection, and is further supported by the fact that the resulting barred galaxies are in good agreement with the {\tt KMeans} classification of centrally elongated systems (\se{classification}).}. 
These bar-like structures are compact -- they have a bar scale $\abar$ that is larger than the effective radii of the inner S\'ersic components, but are all smaller than 1.5 kpc, with the median at 0.87 kpc.

Both the Sérsic indices and the ellipticities indicate that the compact central stellar components are unlikely to be bulges.
The intrinsic 3D shapes and stellar kinematics within the central 1 kpc further support the non-bulge nature of (the majority of) these inner structures. 
Based on the {\tt KMeans} classificaiton in the $q$-$p$-$\vrot/\sigma$ parameter space, 60 per cent of the galaxies are classified as centrally elongated ($p\sim q<0.7$ and $\vrot/\sigma>1.6$), 30 per cent as disky ($p<0.7$, $q>0.7$ and $\vrot/\sigma>1.6$), and only 10 per cent spheroidal ($p\sim q>0.7$ and $\vrot/\sigma<1.6$). 

Taken together, these results show that most massive galaxies at high redshift host distinctive compact central components. 
Multiple independent diagnostics consistently indicate that the majority of these inner components are not bulges, but are instead disky or bar-like structures.

To test the robustness of our results to the numerical effects in the simulation, we repeat the same analysis for three TNG50 runs with different resolution levels, TNG50-1 (the fiducial run used in this work), TNG50-2, and TNG50-3\footnote{TNG50-2 and TNG50-3 have 8 and 64 times lower mass resolutions than TNG50-1. The corresponding softening lengths and star-forming gas cell sizes are larger by factors of 2 and 4.}. We find that the mean values of $\Rein$ in TNG50-1, TNG50-2, and TNG50-3 are 0.25, 0.59, and 1.04 kpc, respectively. The difference between TNG50-2 and TNG50-1 is much smaller than that between TNG50-3 and TNG50-2, indicating that the size measurements are approaching convergence at the TNG50-1 resolution. Meanwhile, the inner S\'ersic indices are consistently low across all three runs, with 72 per cent of the indices below 1.5 in each run. The axis ratios among the three runs also agree well. The mean values of $p$ in TNG50-1, TNG50-2, and TNG50-3 are 0.55, 0.40, and 0.41, respectively, while the corresponding mean values of $q$ are 0.62, 0.64, and 0.72. Therefore, we conclude that our qualitative results are robust to numerical resolution.

\section{Nuclear disks and bars emergence through compaction events}
\label{sec:evolution}

\begin{figure*}[t]
  \centering
  \includegraphics[width=0.87\linewidth]{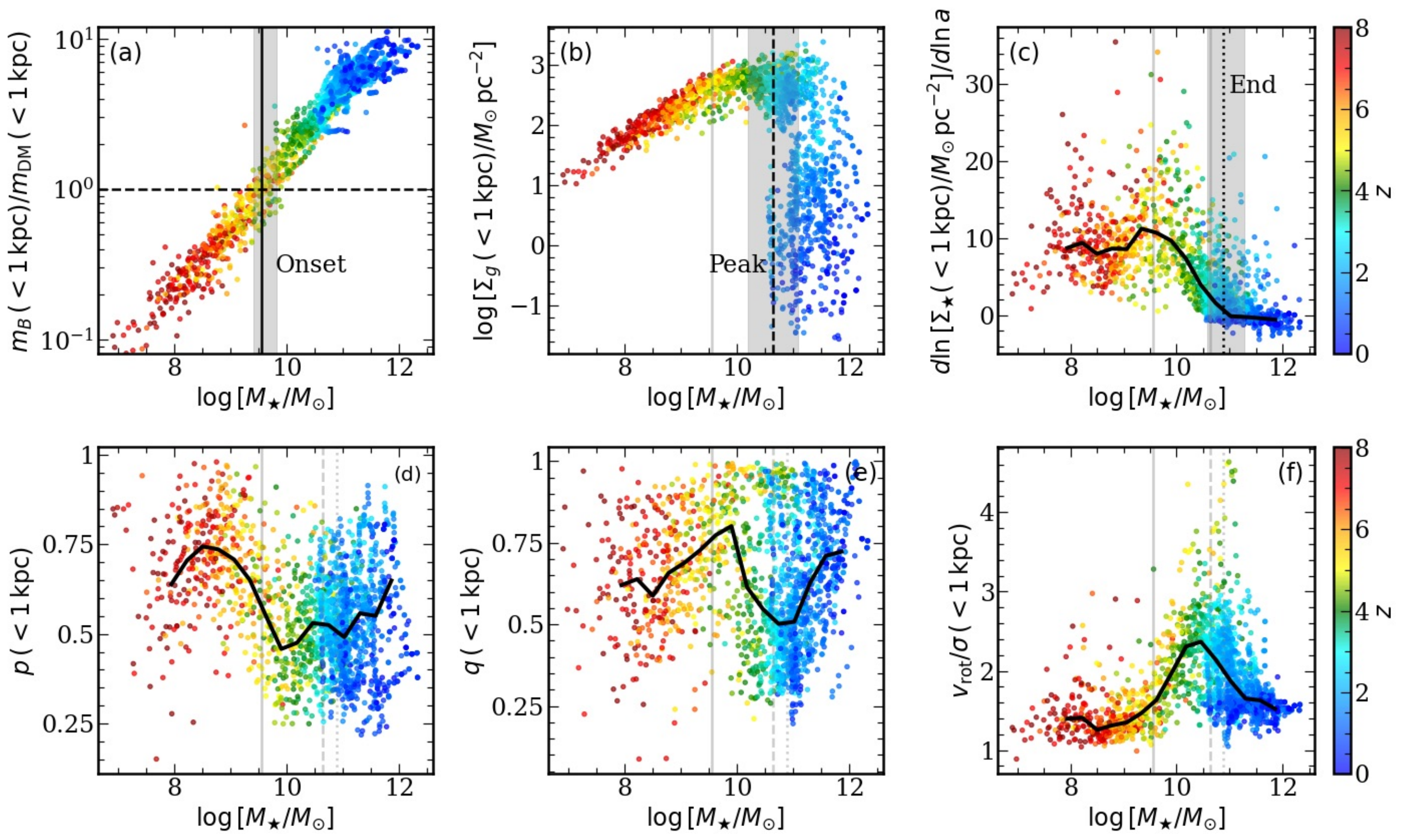} 
  \caption{
{\bf Evolution of the central 1 kpc properties of the 45 massive TNG50 galaxies identified at $z=3$, shown as a function of stellar mass. }
Symbol colors denote redshift, and black curves show the median trends.
(a) Baryon-to-dark-matter mass ratio, $m_{\rm B}/m_{\rm DM}$. The horizontal dashed line marks baryonic dominance ($m_{\rm B}/m_{\rm DM}=1$). 
We define this transition as the {\it onset} of gas-rich compaction, occurring at a characteristic stellar mass of $\Ms\approx10^{9.6}\Msun$. 
(b) Gas surface density, $\Sigmag$. 
The maximum of $\Sigmag$ defines the \textit{peak} of compaction, at $\Ms\approx10^{10.6}\Msun$. 
(c) Growth rate of the central stellar surface density, $d\ln{\Sigmas}/d\ln{a}$, where $a$ is the cosmological scale factor. 
The first time this quantity declines to zero after the peak defines the \textit{end} of compaction, at $\Ms\approx10^{10.9}\Msun$. 
Vertical solid lines mark the median stellar masses of the onset, peak, and end of compaction, while gray bands indicate the corresponding 16th–84th percentile ranges.
\quad
Panels (d)–(f) show the intrinsic axis ratios, $p=c/b$ and $q=b/a$, and the rotation-to-dispersion ratio, $\vrot/\sigma$, measured within the central 1 kpc. 
\quad
Compaction begins at a remarkably well-defined characteristic mass of $\Ms\approx10^{9.6}\Msun$, corresponding to a typical redshift of $z\sim5$ (yellow symbols). 
By the onset of compaction, the central regions are already flattened ($p\simeq0.6$, $q\gtrsim0.75$) and significantly rotation supported ($\vrot/\sigma\gtrsim1.6$). 
As compaction proceeds, the central structures become progressively more elongated while maintaining strong rotational support, indicating that gas-rich compaction builds compact nuclear disks and bars rather than dispersion-dominated spheroidal central components.
}
  \label{fig:compaction}
\end{figure*}

\begin{figure*}[t]
\centering
\includegraphics[width=\linewidth]{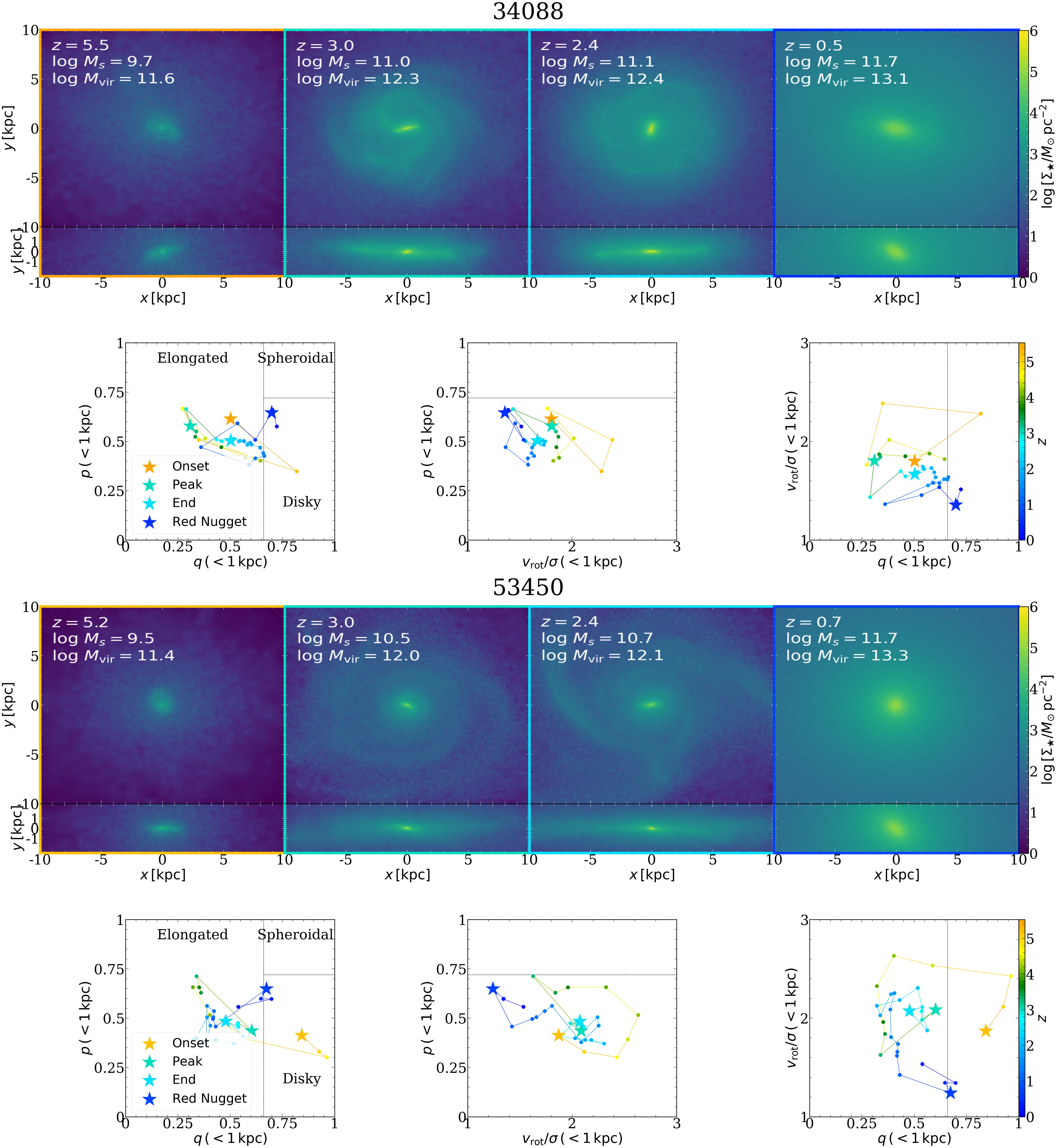}
\caption{
{\bf Two case studies of galaxy evolution in TNG50 (IDs 34088 and 53450), illustrating the emergence of compact nuclear disks and bars during early stages of gas-rich compaction.} 
The logarithmic stellar surface density maps show face-on and edge-on projections of 10 kpc cubic regions at four characteristic epochs, from left to right: the onset of compaction (when the central 1 kpc first becomes baryon dominated), the peak (when the central gas surface density reaches its maximum, at $z\approx3$ for both galaxies), the end (when the central stellar surface density growth rate, $d\ln{\Sigmas}/d\ln{a}$, first declines to zero after the peak, at $z\approx2.4$), and red nugget phase (when the galactic center becomes more spheroidal and dispersion-dominated, after $z\sim1$).
\quad
Below each set of density maps are the corresponding evolution tracks of the galaxies in the three-dimensional parameter space spanned by the intrinsic axis ratios $q=b/a$, $p=c/b$, and the rotation-to-dispersion ratio $\vrot/\sigma$. 
The onset of compaction, the peak, the end and red nugget phase are marked by orange, green, cyan and blue stars, matching the colors of the frames in the density maps.
\quad
Both galaxies enter compaction at $z\gtrsim5$, when their central 1 kpc regions are already significantly rotation supported ($\vrot/\sigma\approx1.8$). 
Galaxy 34088 develops an elongated nucleus immediately, whereas galaxy 53450 evolves from a compact nuclear disk into a bar by the compaction peak. 
In both galaxies, the elongated central structure persists throughout compaction and only transforms into a spheroidal, dispersion-dominated morphology by $z\sim1$.
}
\label{fig:evolution}
\end{figure*}

We trace evolution of the inner structures using the public {\tt SubLink} merger tree\footnote{\url{https://www.tng-project.org/data/docs/specifications/\#sec4a}},
and follow the properties within the central 1 kpc of each galaxy, including the gas surface density $\Sigmag$, the growth rate of the central stellar surface density $d\ln{\Sigmas}/d\ln{a}$ (where $a$ here is the cosmological scale factor instead of the semi-major-axis distance), the baryon-to-dark-matter mass ratio ($\Mb/\Mdm$), the intrinsic axis ratios ($q$ and $p$), and the rotation-to-dispersion ratio ($v_{\text{rot}}/\sigma$), as shown in \Fig{compaction}.
Our goal here is to examine whether the evolution of massive TNG50 galaxies follows the gas-rich compaction scenario, and how the central morphology evolves during compaction.
\subsection{Evidence of compaction in massive TNG50 galaxies}
\label{sec:CompactionEvidence}
Gas-rich compaction refers to a phase in which galaxies undergo central gas condensation due to angular-momentum loss. The infalling gas fuels intense star formation in the galactic center and builds a centrally concentrated stellar core, dubbed as the ``blue nugget''. As the gas is rapidly depleted, star formation declines and the young stellar population passively evolves into a compact quiescent ``red nugget” \citep{Zolotov15, Tacchella16}.

As shown in \Fig{compaction}(a), when gas accumulates toward the center at the beginning of compaction, the central regions evolve from dark matter dominated to baryon dominated, with the baryon-to-dark-matter mass ratio $m_{\rm B}/m_{\rm DM}$ exceeding unity at $\Ms\simeq10^{9.6}\Msun$ (corresponding to a halo virial mass of $\Mv\simeq10^{11.6}\Msun$) which we define as the {\it onset} of compaction\footnote{\citet{Lapiner23} also identify this transition to baryon dominance as a distinct stage preceding the peak of compaction, but adopt a slightly different definition of the onset of compaction.}.
Continued gas inflow drives the central gas surface density $\Sigmag$ to a maximum at $\Ms\simeq10^{10.6}\Msun$ ($\Mv\simeq10^{12.1}\Msun$), marking the {\it peak} of compaction (\Fig{compaction}b). 
As the gas is consumed by star formation and eventually depleted, the growth rate of the central stellar surface density $d\ln{\Sigmas}/d\ln{a}$ declines to nearly zero at $\Ms\simeq10^{10.9}\Msun$ ($\Mv\simeq10^{12.2}\Msun$), signaling the {\it end} of compaction (\Fig{compaction}c). 
Hence, overall, the TNG50 galaxies exhibit the characteristic gas and stellar density evolution patterns expected in the gas-rich compaction scenario \citep{Ceverino14,Ceverino15,Zolotov15,Dubois21}. 
The characteristic mass scales of the onset and peak of compaction for the TNG50 simulations are also broadly consistent with previous studies using different simulations \citep{Lapiner23}.

\subsection{Morphological transformation during compaction}
\label{sec:Transformation}
Compaction is accompanied by significant morphological transformation. 
\Fig{compaction}(d)-(f) demonstrates the evolution in the intrinsic axis ratios ($q$ and $p$) and the rotation-to-dispersion ratio ($v_{\text{rot}}/\sigma$). Following the onset of compaction, the value of $p$ decreases substantially, while $q$ exhibits a similar rapid decline at $z\sim5$. At the same time, the value of $v_{\mathrm{rot}}/\sigma$ increases remarkably. At the peak of compaction, the median intrinsic axis ratios are $q\simeq0.5$ and $p\simeq0.5$ and $\vrot/\sigma$ commonly reaches $2.5$. These suggest that the central 1 kpc rapidly transforms to be flattened and rotation supported as compaction begins, and such characteristics become increasingly pronounced as compaction proceeds. \Fig{classification} further demonstrates the evolutionary picture. At the onset epoch, 60 per cent of the galaxies are classified as disky, 27 per cent as elongated, and only 13 per cent as spheroidal. 
Their central kinematics are likewise dominated by rotation, with 55 per cent of the galaxies having $1<\vrot/\sigma<1.6$ within the central 1 kpc. 
As compaction proceeds, the central structures evolve toward smaller intrinsic axis ratios and stronger rotational support: by $z=3$, the fraction of elongated systems increases to 60 per cent, while 90 per cent of the galaxies have $\vrot/\sigma>1.6$. 

\Fig{evolution} presents two representative examples illustrating the emergence of nuclear disks and bars during gas-rich compaction. 
Both galaxies enter compaction as early as the cosmic dawn of massive galaxy assembly ($z\gtrsim5$), by which time their central 1 kpc regions are already significantly rotation supported, with $\vrot/\sigma\approx1.8$. 
Galaxy 34088 develops an elongated central structure immediately at the onset of compaction, which becomes progressively more prominent toward the compaction peak at $z\approx3$. 
Galaxy 53450 first develops a compact nuclear disk at the onset and subsequently evolves into a barred center by the compaction peak. 
In both galaxies, the elongated central structures persist throughout compaction and remain prominent after compaction ends at $z\approx2.4$. Only at later times ($z\sim1$) do the central regions evolve into more spheroidal, dispersion-dominated systems.

\subsection{Extended disk growth without bulge}
\label{sec:DiskGrowth}
Compaction gives rise to a central stellar core, which subsequently stabilizes newly accreted high-angular-momentum gas and facilitates the formation of an extended disk \citep[e.g.,][]{Dekel20b, Jiang19}. 
This scenario is broadly supported by cosmological hydrodynamical simulations, not only in the {\tt VELA} suite from which it was originally developed \citep{Ceverino14}, but also in the {\tt New-Horizon} simulations \citep{Dubois21}. However, such a stellar core, or the so-called “nugget,” has often been described as a bulge-like or spheroidal structure. Our results suggest that this interpretation is incomplete, as these compact stellar structures exhibit a different morphology of being flattened and rotation-supported. Therefore, the existence of a central stellar bulge is not a necessary prerequisite for the emergence of long-lived disks. Instead, we show that extended disk growth can proceed while the central stellar structure remains rotation-supported and has not yet evolved into a dispersion-dominated spheroid.

\begin{figure}[t]
  \centering
  \includegraphics[width=\linewidth]{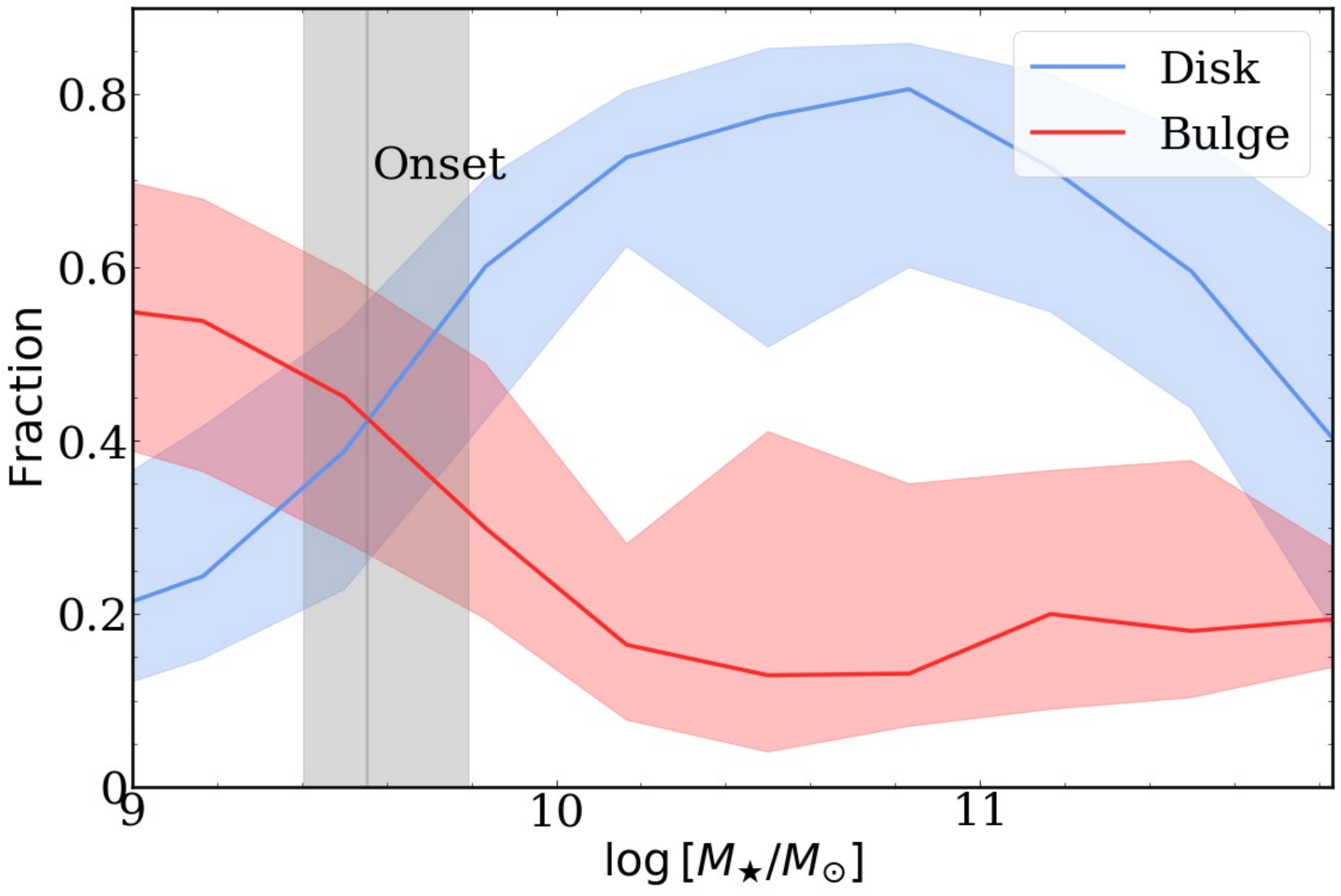} 
  \caption{
{\bf Evolution of the bulge (red) and disk (blue) mass fractions, obtained based on kinematic decomposition.} The vertical gray line marks the onset of compaction. Solid lines represent the median values and shaded regions indicate the $16^\mathrm{th}$--$84^\mathrm{th}$ percentile range. Before compaction, the galaxies are predominantly bulge-dominated. Following the onset of compaction, the disk fraction increases to exceed the bulge fraction. As flattened and rotation-supported stellar cores emerge during compaction, galaxies develop extended disks, with the disk fraction rising to $0.8$, before a dispersion-dominated spheroidal component has formed in the galactic center.
}
  \label{fig:disk}
\end{figure}

\Fig{disk} shows the disk and bulge mass fractions, $f_{\rm disk}$ and $f_{\rm bulge}$, obtained using a kinematic decomposition method \citep{Liang25}, as functions of stellar mass. Before compaction, the galaxies show weak rotational support, with a median disk mass fraction of $\sim0.3$. Around the onset of compaction, the disk fraction starts to increase and surpasses the bulge fraction at $\Ms\simeq10^{9.6}\Msun$. As compaction proceeds and flattened stellar cores emerge, galaxies develop more prominent disks, with the disk fraction eventually reaching $\sim0.8$, before a bulge has formed in the galactic center. 

\section{Conclusion} 
\label{sec:conclusion}

Using the TNG50 cosmological simulation, we have investigated the morphology and evolution of the compact central regions of the complete sample of 45 massive galaxies ($M_\star>10^{10.5}\Msun$) at $z=3$. Our main conclusions are as follows.

\begin{itemize}[leftmargin=*]

\item Distinct compact inner components are ubiquitous in massive galaxies at high redshift. 
More than 90 per cent of the galaxies are better described by a double-Sérsic model than by a single-Sérsic profile, indicating that a separate central component is nearly universal. 
These structures are compact, with typical effective radii of $\lesssim0.5$ kpc, and have low Sérsic indices ($n\sim1$), inconsistent with bulges.

\item Multiple independent diagnostics consistently identify these compact inner components as nuclear disks and bars rather than classical bulges. Double-Sérsic decomposition, isodensity ellipse fitting, intrinsic three-dimensional shapes, and stellar kinematics all indicate flattened or elongated, rotation-supported structures. 
More than half of the sample hosts elongated nuclei, about one third host compact nuclear disks, and only a small minority are consistent with spheroidal morphologies.

\item  These nuclear disks and bars emerge during gas-rich compaction. 
Compaction typically begins at $z\sim5$ and peaks near $z\sim3$, during which the central regions rapidly become flattened and rotation supported. 
Instead of immediately producing the classical “blue nugget” envisioned as a compact spheroid, gas-rich compaction first builds compact nuclear disks and bars that remain in place for a long time before evolving into bulges at much later epochs ($z\lesssim1$). Extended stellar disks grow during this stage, demonstrating that the formation of a classical bulge is not a necessary prerequisite for long-lived disk growth.

\end{itemize}

Overall, this work provides a more comprehensive picture of the galactic centers in the early Universe. The morphological diversity we shown suggests that the central regions encode much richer information than previously assumed. These structures offer a valuable fossil record of disk assembly history, serving as a powerful new test for models of galaxy formation.

\begin{acknowledgments}
We dedicate this work to the memory of Avishai Dekel, whose insight, generosity, and vision continue to inspire our research. 
Avishai played a pioneering role in developing the gas-rich compaction paradigm of high-redshift galaxy evolution, upon which this work builds.
F.J. acknowledges helpful discussions with Aaron Ludlow, and support from the National Natural Science Foundation of China (NSFC; Grant No. 12473007) and the China Manned Space Program (Grant No. CMS-CSST-2025-A03). L.C.H. acknowledges support from the National Natural Science Foundation of China (Grant No. 12233001) and the China Manned Space Program (Grant No. CMS-CSST-2025-A09). Y.P. acknowledges support from the National Natural Science Foundation of China (NSFC) under grant Nos. 12125301, 12192220, and 12192222, and from the New Cornerstone Science Foundation through the XPLORER PRIZE.
\end{acknowledgments}

\software{lmfit \citep{Newville25}; astropy \citep{Bradley25}; scikit-learn \citep{sklearn25}}


\bibliography{ref.bib}

@ARTICLE{Aguerri09,
       author = {{Aguerri}, J.~A.~L. and {M{\'e}ndez-Abreu}, J. and {Corsini}, E.~M.},
        title = "{The population of barred galaxies in the local universe. I. Detection and characterisation of bars}",
      journal = {\aap},
         year = 2009,
        month = feb,
       volume = {495},
       number = {2},
        pages = {491-504},
          doi = {10.1051/0004-6361:200810931},
archivePrefix = {arXiv},
       eprint = {0901.2346},
 primaryClass = {astro-ph.GA},
       adsurl = {https://ui.adsabs.harvard.edu/abs/2009A&A...495..491A}
}

@ARTICLE{Athanassoula02,
       author = {{Athanassoula}, E. and {Misiriotis}, A.},
        title = "{Morphology, photometry and kinematics of N -body bars - I. Three models with different halo central concentrations}",
      journal = {\mnras},
         year = 2002,
        month = feb,
       volume = {330},
       number = {1},
        pages = {35-52},
          doi = {10.1046/j.1365-8711.2002.05028.x},
archivePrefix = {arXiv},
       eprint = {astro-ph/0111449},
 primaryClass = {astro-ph},
       adsurl = {https://ui.adsabs.harvard.edu/abs/2002MNRAS.330...35A}
}

@ARTICLE{Bland-Hawthorn16,
       author = {{Bland-Hawthorn}, Joss and {Gerhard}, Ortwin},
        title = "{The Galaxy in Context: Structural, Kinematic, and Integrated Properties}",
      journal = {\araa},
         year = 2016,
        month = sep,
       volume = {54},
        pages = {529-596},
          doi = {10.1146/annurev-astro-081915-023441},
archivePrefix = {arXiv},
       eprint = {1602.07702},
 primaryClass = {astro-ph.GA},
       adsurl = {https://ui.adsabs.harvard.edu/abs/2016ARA&A..54..529B}
}

@ARTICLE{Boker02,
       author = {{B{\"o}ker}, Torsten and {Laine}, Seppo and {van der Marel}, Roeland P. and {Sarzi}, Marc and {Rix}, Hans-Walter and {Ho}, Luis C. and {Shields}, Joseph C.},
        title = "{A Hubble Space Telescope Census of Nuclear Star Clusters in Late-Type Spiral Galaxies. I. Observations and Image Analysis}",
      journal = {\aj},
         year = 2002,
        month = mar,
       volume = {123},
       number = {3},
        pages = {1389-1410},
          doi = {10.1086/339025},
archivePrefix = {arXiv},
       eprint = {astro-ph/0112086},
 primaryClass = {astro-ph},
       adsurl = {https://ui.adsabs.harvard.edu/abs/2002AJ....123.1389B}
}

@software{Bradley25,
  author       = {Larry Bradley and
                  Brigitta Sip{\H o}cz and
                  Thomas Robitaille and
                  Erik Tollerud and
                  Z\`e Vin{\'{\i}}cius and
                  Christoph Deil and
                  Kyle Barbary and
                  Tom J Wilson and
                  Ivo Busko and
                  Axel Donath and
                  Hans Moritz G{\"u}nther and
                  Mihai Cara and
                  P. L. Lim and
                  Sebastian Me{\ss}linger and
                  Zach Burnett and
                  Simon Conseil and
                  Michael Droettboom and
                  Azalee Bostroem and
                  E. M. Bray and
                  Lars Andersen Bratholm and
                  William Jamieson and
                  Adam Ginsburg and
                  Geert Barentsen and
                  Matt Craig and
                  Sergio Pascual and
                  Shivangee Rathi and
                  Marshall Perrin and
                  Brett M. Morris},
  title        = {astropy/photutils: 2.2.0},
  month        = feb,
  year         = 2025,
  publisher    = {Zenodo},
  version      = {2.2.0},
  doi          = {10.5281/zenodo.14889440},
  url          = {https://doi.org/10.5281/zenodo.14889440},
  swhid        = {swh:1:dir:11159107f27a28985192ed1118b1f2055709d093
                   ;origin=https://doi.org/10.5281/zenodo.596036;visi
                   t=swh:1:snp:ae8c4a55d349d43e53cfe9ce92e678fcfe840f
                   3b;anchor=swh:1:rel:0117f67e8888adcdfc85308287dd9c
                   854b466389;path=astropy-photutils-ffb96c5
                  },
}

@ARTICLE{Ceverino14,
       author = {{Ceverino}, Daniel and {Klypin}, Anatoly and {Klimek}, Elizabeth S. and {Trujillo-Gomez}, Sebastian and {Churchill}, Christopher W. and {Primack}, Joel and {Dekel}, Avishai},
        title = "{Radiative feedback and the low efficiency of galaxy formation in low-mass haloes at high redshift}",
      journal = {\mnras},
         year = 2014,
        month = aug,
       volume = {442},
       number = {2},
        pages = {1545-1559},
          doi = {10.1093/mnras/stu956},
archivePrefix = {arXiv},
       eprint = {1307.0943},
 primaryClass = {astro-ph.CO},
       adsurl = {https://ui.adsabs.harvard.edu/abs/2014MNRAS.442.1545C}
}

@ARTICLE{Ceverino15,
       author = {{Ceverino}, Daniel and {Dekel}, Avishai and {Tweed}, Dylan and {Primack}, Joel},
        title = "{Early formation of massive, compact, spheroidal galaxies with classical profiles by violent disc instability or mergers}",
      journal = {\mnras},
         year = 2015,
        month = mar,
       volume = {447},
       number = {4},
        pages = {3291-3310},
          doi = {10.1093/mnras/stu2694},
archivePrefix = {arXiv},
       eprint = {1409.2622},
 primaryClass = {astro-ph.GA},
       adsurl = {https://ui.adsabs.harvard.edu/abs/2015MNRAS.447.3291C}
}

@ARTICLE{Ciotti99,
       author = {{Ciotti}, L. and {Bertin}, G.},
        title = "{Analytical properties of the R$^{1/m}$ law}",
      journal = {\aap},
         year = 1999,
        month = dec,
       volume = {352},
        pages = {447-451},
          doi = {10.48550/arXiv.astro-ph/9911078},
archivePrefix = {arXiv},
       eprint = {astro-ph/9911078},
 primaryClass = {astro-ph},
       adsurl = {https://ui.adsabs.harvard.edu/abs/1999A&A...352..447C}
}

@article{Conselice14,
   title={The Evolution of Galaxy Structure Over Cosmic Time},
   volume={52},
   ISSN={1545-4282},
   url={http://dx.doi.org/10.1146/annurev-astro-081913-040037},
   DOI={10.1146/annurev-astro-081913-040037},
   number={1},
   journal={Annual Review of Astronomy and Astrophysics},
   publisher={Annual Reviews},
   author={Conselice, Christopher J.},
   year={2014},
   month=Aug, pages={291–337} }

@ARTICLE{Costantin23,
       author = {{Costantin}, Luca and {P{\'e}rez-Gonz{\'a}lez}, Pablo G. and {Vega-Ferrero}, Jes{\'u}s and {Huertas-Company}, Marc and {Bisigello}, Laura and {Buitrago}, Fernando and {Bagley}, Micaela B. and {Cleri}, Nikko J. and {Cooper}, Michael C. and {Finkelstein}, Steven L. and {Holwerda}, Benne W. and {Kartaltepe}, Jeyhan S. and {Koekemoer}, Anton M. and {Nelson}, Dylan and {Papovich}, Casey and {Pillepich}, Annalisa and {Pirzkal}, Nor and {Tacchella}, Sandro and {Yung}, L.~Y. Aaron},
        title = "{Expectations of the Size Evolution of Massive Galaxies at 3 {\ensuremath{\leq}} z {\ensuremath{\leq}} 6 from the TNG50 Simulation: The CEERS/JWST View}",
      journal = {\apj},
         year = 2023,
        month = apr,
       volume = {946},
       number = {2},
          eid = {71},
        pages = {71},
          doi = {10.3847/1538-4357/acb926},
archivePrefix = {arXiv},
       eprint = {2208.00007},
 primaryClass = {astro-ph.GA},
       adsurl = {https://ui.adsabs.harvard.edu/abs/2023ApJ...946...71C}
}

@ARTICLE{Dekel14,
       author = {{Dekel}, A. and {Burkert}, A.},
        title = "{Wet disc contraction to galactic blue nuggets and quenching to red nuggets}",
      journal = {\mnras},
         year = 2014,
        month = feb,
       volume = {438},
       number = {2},
        pages = {1870-1879},
          doi = {10.1093/mnras/stt2331},
archivePrefix = {arXiv},
       eprint = {1310.1074},
 primaryClass = {astro-ph.CO},
       adsurl = {https://ui.adsabs.harvard.edu/abs/2014MNRAS.438.1870D}
}

@ARTICLE{Dekel20b,
       author = {{Dekel}, Avishai and {Lapiner}, Sharon and {Ginzburg}, Omri and {Freundlich}, Jonathan and {Jiang}, Fangzhou and {Finish}, Bar and {Kretschmer}, Michael and {Lin}, Doug and {Ceverino}, Daniel and {Primack}, Joel and et al.},
        title = "{Origin of star-forming rings around massive centres in massive galaxies at z < 4}",
      journal = {\mnras},
         year = 2020,
        month = aug,
       volume = {496},
       number = {4},
        pages = {5372-5398},
          doi = {10.1093/mnras/staa1713},
archivePrefix = {arXiv},
       eprint = {2003.08984},
 primaryClass = {astro-ph.GA},
       adsurl = {https://ui.adsabs.harvard.edu/abs/2020MNRAS.496.5372D}
}

@ARTICLE{Dubois21,
       author = {{Dubois}, Yohan and {Beckmann}, Ricarda and {Bournaud}, Fr{\'e}d{\'e}ric and {Choi}, Hoseung and {Devriendt}, Julien and {Jackson}, Ryan and {Kaviraj}, Sugata and {Kimm}, Taysun and {Kraljic}, Katarina and {Laigle}, Clotilde and et al.},
        title = "{Introducing the NEWHORIZON simulation: Galaxy properties with resolved internal dynamics across cosmic time}",
      journal = {\aap},
         year = 2021,
        month = jul,
       volume = {651},
          eid = {A109},
        pages = {A109},
          doi = {10.1051/0004-6361/202039429},
archivePrefix = {arXiv},
       eprint = {2009.10578},
 primaryClass = {astro-ph.GA},
       adsurl = {https://ui.adsabs.harvard.edu/abs/2021A&A...651A.109D}
}

@ARTICLE{Erwin24,
       author = {{Erwin}, Peter},
        title = "{The frequency and sizes of inner bars and nuclear rings in barred galaxies and their dependence on galaxy properties}",
      journal = {\mnras},
         year = 2024,
        month = feb,
       volume = {528},
       number = {2},
        pages = {3613-3628},
          doi = {10.1093/mnras/stad3944},
archivePrefix = {arXiv},
       eprint = {2312.12893},
 primaryClass = {astro-ph.GA},
       adsurl = {https://ui.adsabs.harvard.edu/abs/2024MNRAS.528.3613E}
}

@ARTICLE{Ferreira23,
       author = {{Ferreira}, Leonardo and {Conselice}, Christopher J. and {Sazonova}, Elizaveta and {Ferrari}, Fabricio and {Caruana}, Joseph and {Tohill}, Cl{\'a}r-Br{\'\i}d and {Lucatelli}, Geferson and {Adams}, Nathan and {Irodotou}, Dimitrios and {Marshall}, Madeline A. and {Roper}, Will J. and {Lovell}, Christopher C. and {Verma}, Aprajita and {Austin}, Duncan and {Trussler}, James and {Wilkins}, Stephen M.},
        title = "{The JWST Hubble Sequence: The Rest-frame Optical Evolution of Galaxy Structure at 1.5 < z < 6.5}",
      journal = {\apj},
         year = 2023,
        month = oct,
       volume = {955},
       number = {2},
          eid = {94},
        pages = {94},
          doi = {10.3847/1538-4357/acec76},
archivePrefix = {arXiv},
       eprint = {2210.01110},
 primaryClass = {astro-ph.GA},
       adsurl = {https://ui.adsabs.harvard.edu/abs/2023ApJ...955...94F}
}

@ARTICLE{ForouharMoreno26,
       author = {{Forouhar Moreno}, Victor J. and {Schaye}, Joop and {Schaller}, Matthieu and {Ludlow}, Aaron and {McGibbon}, Robert J. and {Ben{\'\i}tez-Llambay}, Alejandro and {Chaikin}, Evgenii and {Frenk}, Carlos S. and {Hu{\v{s}}ko}, Filip and {Ploeckinger}, Sylvia and {Richings}, Alexander J. and {Trayford}, James W.},
        title = "{The morphologies of present-day galaxies in the COLIBRE simulations}",
      journal = {arXiv e-prints},
         year = 2026,
        month = apr,
          eid = {arXiv:2604.03503},
        pages = {arXiv:2604.03503},
          doi = {10.48550/arXiv.2604.03503},
archivePrefix = {arXiv},
       eprint = {2604.03503},
 primaryClass = {astro-ph.GA},
       adsurl = {https://ui.adsabs.harvard.edu/abs/2026arXiv260403503F}
}

@ARTICLE{Garland26,
       author = {{Garland}, I.~L. and {Best}, H. and {Fortson}, L.~F. and {G{\'e}ron}, T. and {Lintott}, C.~J. and {O'Ryan}, D. and {Simmons}, B.~D. and {Smethurst}, R.~J. and {Viskotov{\'a}}, M. and {Walmsley}, M. and et al.},
        title = "{The complex relationships between active galactic nuclei, bars, and bulges}",
      journal = {\aap},
         year = 2026,
        month = may,
       volume = {709},
          eid = {A48},
        pages = {A48},
          doi = {10.1051/0004-6361/202557755},
archivePrefix = {arXiv},
       eprint = {2603.28208},
 primaryClass = {astro-ph.GA},
       adsurl = {https://ui.adsabs.harvard.edu/abs/2026A&A...709A..48G}
}

@ARTICLE{Gong25,
       author = {{Gong}, Jun-Yu and {Lin}, Weipeng and {Tang}, Lin and {Lan}, Yanyao},
        title = "{Mock Observations: Morphological Analysis of Galaxies in TNG100 Simulations}",
      journal = {\apjs},
         year = 2025,
        month = jul,
       volume = {279},
       number = {1},
          eid = {16},
        pages = {16},
          doi = {10.3847/1538-4365/add5e4},
archivePrefix = {arXiv},
       eprint = {2504.18042},
 primaryClass = {astro-ph.GA},
       adsurl = {https://ui.adsabs.harvard.edu/abs/2025ApJS..279...16G}
}

@ARTICLE{Huang13,
       author = {{Huang}, Song and {Ho}, Luis C. and {Peng}, Chien Y. and {Li}, Zhao-Yu and {Barth}, Aaron J.},
        title = "{The Carnegie-Irvine Galaxy Survey. III. The Three-component Structure of Nearby Elliptical Galaxies}",
      journal = {\apj},
         year = 2013,
        month = mar,
       volume = {766},
       number = {1},
          eid = {47},
        pages = {47},
          doi = {10.1088/0004-637X/766/1/47},
archivePrefix = {arXiv},
       eprint = {1212.2639},
 primaryClass = {astro-ph.CO},
       adsurl = {https://ui.adsabs.harvard.edu/abs/2013ApJ...766...47H}
}

@ARTICLE{Huertas-Company24,
       author = {{Huertas-Company}, M. and {Iyer}, K.~G. and {Angeloudi}, E. and {Bagley}, M.~B. and {Finkelstein}, S.~L. and {Kartaltepe}, J. and {McGrath}, E.~J. and {Sarmiento}, R. and {Vega-Ferrero}, J. and {Arrabal Haro}, P. and {Behroozi}, P. and {Buitrago}, F. and {Cheng}, Y. and {Costantin}, L. and {Dekel}, A. and {Dickinson}, M. and {Elbaz}, D. and {Grogin}, N.~A. and {Hathi}, N.~P. and {Holwerda}, B.~W. and {Koekemoer}, A.~M. and {Lucas}, R.~A. and {Papovich}, C. and {P{\'e}rez-Gonz{\'a}lez}, P.~G. and {Pirzkal}, N. and {Seill{\'e}}, L.-M. and {de la Vega}, A. and {Wuyts}, S. and {Yang}, G. and {Yung}, L.~Y.~A.},
        title = "{Galaxy morphology from z {\ensuremath{\sim}} 6 through the lens of JWST}",
      journal = {\aap},
         year = 2024,
        month = may,
       volume = {685},
          eid = {A48},
        pages = {A48},
          doi = {10.1051/0004-6361/202346800},
archivePrefix = {arXiv},
       eprint = {2305.02478},
 primaryClass = {astro-ph.GA},
       adsurl = {https://ui.adsabs.harvard.edu/abs/2024A&A...685A..48H}
}

@ARTICLE{Huertas-Company25,
       author = {{Huertas-Company}, M. and {Shuntov}, M. and {Dong}, Y. and {Walmsley}, M. and {Ilbert}, O. and {McCracken}, H.~J. and {Akins}, H.~B. and {Allen}, N. and {Casey}, C.~M. and {Costantin}, L. and {Daddi}, E. and {Dekel}, A. and {Franco}, M. and {Garland}, I.~L. and {G{\'e}ron}, T. and {Gozaliasl}, G. and {Hirschmann}, M. and {Kartaltepe}, J.~S. and {Koekemoer}, A.~M. and {Lintott}, C. and {Liu}, D. and {Lucas}, R. and {Masters}, K. and {Pacucci}, F. and {Paquereau}, L. and {P{\'e}rez-Gonz{\'a}lez}, P.~G. and {Rhodes}, J.~D. and {Robertson}, B.~E. and {Simmons}, B. and {Smethurst}, R. and {Toft}, S. and {Yang}, L.},
        title = "{COSMOS-Web: The emergence of the Hubble sequence}",
      journal = {\aap},
         year = 2025,
        month = dec,
       volume = {704},
          eid = {A94},
        pages = {A94},
          doi = {10.1051/0004-6361/202553782},
archivePrefix = {arXiv},
       eprint = {2502.03532},
 primaryClass = {astro-ph.GA},
       adsurl = {https://ui.adsabs.harvard.edu/abs/2025A&A...704A..94H}
}

@ARTICLE{Jiang19,
       author = {{Jiang}, Fangzhou and {Dekel}, Avishai and {Kneller}, Omer and {Lapiner}, Sharon and {Ceverino}, Daniel and {Primack}, Joel R. and {Faber}, Sandra M. and {Macci{\`o}}, Andrea V. and {Dutton}, Aaron A. and {Genel}, Shy and et al.},
        title = "{Is the dark-matter halo spin a predictor of galaxy spin and size?}",
      journal = {\mnras},
         year = 2019,
        month = oct,
       volume = {488},
       number = {4},
        pages = {4801-4815},
          doi = {10.1093/mnras/stz1952},
archivePrefix = {arXiv},
       eprint = {1804.07306},
 primaryClass = {astro-ph.GA},
       adsurl = {https://ui.adsabs.harvard.edu/abs/2019MNRAS.488.4801J}
}

@ARTICLE{Jiang25,
       author = {{Jiang}, Fangzhou and {Liang}, Jinning and {Jin}, Bingcheng and {Gao}, Zeyu and {Wang}, Weichen and {Cantalupo}, Sebastiano and {Shen}, Xuejian and {Ho}, Luis C. and {Peng}, Yingjie and {Wang}, Jing},
        title = "{Formation and Environmental Context of Giant Bulgeless Disk Galaxies in the Early Universe: Insights from Cosmological Simulations}",
      journal = {arXiv e-prints},
         year = 2025,
        month = apr,
          eid = {arXiv:2504.01070},
        pages = {arXiv:2504.01070},
          doi = {10.48550/arXiv.2504.01070},
archivePrefix = {arXiv},
       eprint = {2504.01070},
 primaryClass = {astro-ph.GA},
       adsurl = {https://ui.adsabs.harvard.edu/abs/2025arXiv250401070J}
}

@article{Kartaltepe23,
   title={CEERS Key Paper. III. The Diversity of Galaxy Structure and Morphology at z = 3–9 with JWST},
   volume={946},
   ISSN={2041-8213},
   url={http://dx.doi.org/10.3847/2041-8213/acad01},
   DOI={10.3847/2041-8213/acad01},
   number={1},
   journal={The Astrophysical Journal Letters},
   publisher={American Astronomical Society},
   author={Kartaltepe, Jeyhan S. and Rose, Caitlin and Vanderhoof, Brittany N. and McGrath, Elizabeth J. and Costantin, Luca and Cox, Isabella G. and Yung, L. Y. Aaron and Kocevski, Dale D. and Wuyts, Stijn and Ferguson, Henry C. and Bagley, Micaela B. and Finkelstein, Steven L. and Amorín, Ricardo O. and Andrews, Brett H. and Arrabal Haro, Pablo and Backhaus, Bren E. and Behroozi, Peter and Bisigello, Laura and Calabrò, Antonello and Casey, Caitlin M. and Coogan, Rosemary T. and Cooper, M. C. and Croton, Darren and de la Vega, Alexander and Dickinson, Mark and Fontana, Adriano and Franco, Maximilien and Grazian, Andrea and Grogin, Norman A. and Hathi, Nimish P. and Holwerda, Benne W. and Huertas-Company, Marc and Iyer, Kartheik G. and Jogee, Shardha and Jung, Intae and Kewley, Lisa J. and Kirkpatrick, Allison and Koekemoer, Anton M. and Liu, James and Lotz, Jennifer M. and Lucas, Ray A. and Newman, Jeffrey A. and Pacifici, Camilla and Pandya, Viraj and Papovich, Casey and Pentericci, Laura and Pérez-González, Pablo G. and Petersen, Jayse and Pirzkal, Nor and Rafelski, Marc and Ravindranath, Swara and Simons, Raymond C. and Snyder, Gregory F. and Somerville, Rachel S. and Stanway, Elizabeth R. and Straughn, Amber N. and Tacchella, Sandro and Trump, Jonathan R. and Vega-Ferrero, Jesús and Wilkins, Stephen M. and Yang, Guang and Zavala, Jorge A.},
   year={2023},
   month=Mar, pages={L15} }

@ARTICLE{Kim21,
       author = {{Kim}, Taehyun and {Athanassoula}, E. and {Sheth}, Kartik and {Bosma}, Albert and {Park}, Myeong-Gu and {Lee}, Yun Hee and {Ann}, Hong Bae},
        title = "{Cosmic Evolution of Barred Galaxies up to z   0.84}",
      journal = {\apj},
         year = 2021,
        month = dec,
       volume = {922},
       number = {2},
          eid = {196},
        pages = {196},
          doi = {10.3847/1538-4357/ac2300},
archivePrefix = {arXiv},
       eprint = {2109.03420},
 primaryClass = {astro-ph.GA},
       adsurl = {https://ui.adsabs.harvard.edu/abs/2021ApJ...922..196K}
}

@article{Kormendy04,
   title={Secular Evolution and the Formation of Pseudobulges in Disk Galaxies},
   volume={42},
   ISSN={1545-4282},
   url={http://dx.doi.org/10.1146/annurev.astro.42.053102.134024},
   DOI={10.1146/annurev.astro.42.053102.134024},
   number={1},
   journal={Annual Review of Astronomy and Astrophysics},
   publisher={Annual Reviews},
   author={Kormendy, John and Kennicutt, Robert C.},
   year={2004},
   month=Sept, pages={603–683} }

@article{Kormendy10,
   title={BULGELESS GIANT GALAXIES CHALLENGE OUR PICTURE OF GALAXY FORMATION BY HIERARCHICAL CLUSTERING,},
   volume={723},
   ISSN={1538-4357},
   url={http://dx.doi.org/10.1088/0004-637X/723/1/54},
   DOI={10.1088/0004-637x/723/1/54},
   number={1},
   journal={The Astrophysical Journal},
   publisher={American Astronomical Society},
   author={Kormendy, John and Drory, Niv and Bender, Ralf and Cornell, Mark E.},
   year={2010},
   month=Oct, pages={54–80} }

@article{Lapiner23,
   title={Wet compaction to a blue nugget: a critical phase in galaxy evolution},
   volume={522},
   ISSN={1365-2966},
   url={http://dx.doi.org/10.1093/mnras/stad1263},
   DOI={10.1093/mnras/stad1263},
   number={3},
   journal={Monthly Notices of the Royal Astronomical Society},
   publisher={Oxford University Press (OUP)},
   author={Lapiner, Sharon and Dekel, Avishai and Freundlich, Jonathan and Ginzburg, Omri and Jiang, Fangzhou and Kretschmer, Michael and Tacchella, Sandro and Ceverino, Daniel and Primack, Joel},
   year={2023},
   month=apr, pages={4515–4547} }

@ARTICLE{LeConte24,
       author = {{Le Conte}, Zoe A. and {Gadotti}, Dimitri A. and {Ferreira}, Leonardo and {Conselice}, Christopher J. and {de S{\'a}-Freitas}, Camila and {Kim}, Taehyun and {Neumann}, Justus and {Fragkoudi}, Francesca and {Athanassoula}, E. and {Adams}, Nathan J.},
        title = "{A JWST investigation into the bar fraction at redshifts 1 {\ensuremath{\leq}} z {\ensuremath{\leq}} 3}",
      journal = {\mnras},
         year = 2024,
        month = may,
       volume = {530},
       number = {2},
        pages = {1984-2000},
          doi = {10.1093/mnras/stae921},
archivePrefix = {arXiv},
       eprint = {2309.10038},
 primaryClass = {astro-ph.GA},
       adsurl = {https://ui.adsabs.harvard.edu/abs/2024MNRAS.530.1984L}
}

@ARTICLE{LeConte26,
       author = {{Le Conte}, Zoe A. and {Gadotti}, Dimitri A. and {Harvey}, Thomas and {Ferreira}, Leonardo and {Conselice}, Christopher J. and {Kim}, Taehyun and {de S{\'a}-Freitas}, Camila and {Fragkoudi}, Francesca and {Neumann}, Justus and {Athanassoula}, E.},
        title = "{A nuclear disc at Cosmic Noon: evidence of early bar-driven galaxy evolution}",
      journal = {\mnras},
         year = 2026,
        month = aug,
       volume = {550},
       number = {2},
          eid = {stag1122},
        pages = {stag1122},
          doi = {10.1093/mnras/stag1122},
archivePrefix = {arXiv},
       eprint = {2601.18871},
 primaryClass = {astro-ph.GA},
       adsurl = {https://ui.adsabs.harvard.edu/abs/2026MNRAS.550g1122L}
}

@ARTICLE{Lee24,
       author = {{Lee}, Jeong Hwan and {Park}, Changbom and {Hwang}, Ho Seong and {Kwon}, Minseong},
        title = "{Morphology of Galaxies in JWST Fields: Initial Distribution and Evolution of Galaxy Morphology}",
      journal = {\apj},
         year = 2024,
        month = may,
       volume = {966},
       number = {1},
          eid = {113},
        pages = {113},
          doi = {10.3847/1538-4357/ad3448},
archivePrefix = {arXiv},
       eprint = {2312.04899},
 primaryClass = {astro-ph.GA},
       adsurl = {https://ui.adsabs.harvard.edu/abs/2024ApJ...966..113L}
}

@ARTICLE{Liang25,
       author = {{Liang}, Jinning and {Jiang}, Fangzhou and {Mo}, Houjun and {Benson}, Andrew and {Dekel}, Avishai and {Tavron}, Noa and {Hopkins}, Philip F. and {Ho}, Luis C.},
        title = "{Connection between galaxy morphology and dark-matter halo structure I: a running threshold for thin discs and size predictors from the dark sector}",
      journal = {\mnras},
         year = 2025,
        month = aug,
       volume = {541},
       number = {3},
        pages = {2304-2323},
          doi = {10.1093/mnras/staf947},
archivePrefix = {arXiv},
       eprint = {2403.14749},
 primaryClass = {astro-ph.GA},
       adsurl = {https://ui.adsabs.harvard.edu/abs/2025MNRAS.541.2304L}
}

@misc{Lyu26,
      title={On the Detectability and Measurement of Galactic Bars as a Function of Redshift}, 
      author={Yi-Xiao Lyu and Luis C. Ho and Zhao-Yu Li and Ming-Yang Zhuang},
      year={2026},
      eprint={2609.23501},
      archivePrefix={arXiv},
      primaryClass={astro-ph.GA},
      url={https://arxiv.org/abs/2609.23501}, 
}

@ARTICLE{Nelson23,
       author = {{Nelson}, Erica J. and {Suess}, Katherine A. and {Bezanson}, Rachel and {Price}, Sedona H. and {van Dokkum}, Pieter and {Leja}, Joel and {Wang}, Bingjie and {Whitaker}, Katherine E. and {Labb{\'e}}, Ivo and {Barrufet}, Laia and {Brammer}, Gabriel and {Eisenstein}, Daniel J. and {Gibson}, Justus and {Hartley}, Abigail I. and {Johnson}, Benjamin D. and {Heintz}, Kasper E. and {Mathews}, Elijah and {Miller}, Tim B. and {Oesch}, Pascal A. and {Sandles}, Lester and {Setton}, David J. and {Speagle}, Joshua S. and {Tacchella}, Sandro and {Tadaki}, Ken-ichi and {{\"U}bler}, Hannah and {Weaver}, John. R.},
        title = "{JWST Reveals a Population of Ultrared, Flattened Galaxies at 2 {\ensuremath{\lesssim}} z {\ensuremath{\lesssim}} 6 Previously Missed by HST}",
      journal = {\apjl},
         year = 2023,
        month = may,
       volume = {948},
       number = {2},
          eid = {L18},
        pages = {L18},
          doi = {10.3847/2041-8213/acc1e1},
archivePrefix = {arXiv},
       eprint = {2208.01630},
 primaryClass = {astro-ph.GA},
       adsurl = {https://ui.adsabs.harvard.edu/abs/2023ApJ...948L..18N}
}

@software{Newville25,
  author  = {Newville, M. and Otten, R. and Nelson, A. and
             Stensitzki, T. and Ingargiola, A. and Allan, D. and
             Fox, A. and Carter, F. and Rawlik, M.},
  title   = {{LMFIT}: Non-Linear Least-Squares Minimization and Curve-Fitting for Python},
  year    = {2025},
  version = {1.3.4},
  doi     = {10.5281/zenodo.16175987},
  url     = {https://doi.org/10.5281/zenodo.16175987},
  publisher = {Zenodo},
}

@ARTICLE{Pandya24,
       author = {{Pandya}, Viraj and {Zhang}, Haowen and {Huertas-Company}, Marc and {Iyer}, Kartheik G. and {McGrath}, Elizabeth and {Barro}, Guillermo and {Finkelstein}, Steven L. and {K{\"u}mmel}, Martin and {Hartley}, William G. and {Ferguson}, Henry C. and {Kartaltepe}, Jeyhan S. and {Primack}, Joel and {Dekel}, Avishai and {Faber}, Sandra M. and {Koo}, David C. and {Bryan}, Greg L. and {Somerville}, Rachel S. and {Amor{\'\i}n}, Ricardo O. and {Arrabal Haro}, Pablo and {Bagley}, Micaela B. and {Bell}, Eric F. and {Bertin}, Emmanuel and {Costantin}, Luca and {Dav{\'e}}, Romeel and {Dickinson}, Mark and {Feldmann}, Robert and {Fontana}, Adriano and {Gavazzi}, Raphael and {Giavalisco}, Mauro and {Grazian}, Andrea and {Grogin}, Norman A. and {Guo}, Yuchen and {Hahn}, ChangHoon and {Holwerda}, Benne W. and {Kewley}, Lisa J. and {Kirkpatrick}, Allison and {Kocevski}, Dale D. and {Koekemoer}, Anton M. and {Lotz}, Jennifer M. and {Lucas}, Ray A. and {Papovich}, Casey and {Pentericci}, Laura and {P{\'e}rez-Gonz{\'a}lez}, Pablo G. and {Pirzkal}, Nor and {Ravindranath}, Swara and {Rose}, Caitlin and {Schefer}, Marc and {Simons}, Raymond C. and {Straughn}, Amber N. and {Tacchella}, Sandro and {Trump}, Jonathan R. and {de la Vega}, Alexander and {Wilkins}, Stephen M. and {Wuyts}, Stijn and {Yang}, Guang and {Yung}, L.~Y. Aaron},
        title = "{Galaxies Going Bananas: Inferring the 3D Geometry of High-redshift Galaxies with JWST-CEERS}",
      journal = {\apj},
         year = 2024,
        month = mar,
       volume = {963},
       number = {1},
          eid = {54},
        pages = {54},
          doi = {10.3847/1538-4357/ad1a13},
archivePrefix = {arXiv},
       eprint = {2310.15232},
 primaryClass = {astro-ph.GA},
       adsurl = {https://ui.adsabs.harvard.edu/abs/2024ApJ...963...54P}
}

@article{Pillepich19,
    author = "Pillepich, Annalisa and others",
    title = "{First results from the TNG50 simulation: the evolution of stellar and gaseous discs across cosmic time}",
    eprint = "1902.05553",
    archivePrefix = "arXiv",
    primaryClass = "astro-ph.GA",
    doi = "10.1093/mnras/stz2338",
    journal = "Mon. Not. Roy. Astron. Soc.",
    volume = "490",
    number = "3",
    pages = "3196--3233",
    year = "2019"
}

@ARTICLE{Quadri26,
       author = {{Quadri}, G. and {Cantalupo}, S. and {Bacchini}, C. and {Pensabene}, A. and {Lupi}, A. and {Pezzulli}, G. and {Wang}, W. and {Galbiati}, M. and {Lazeyras}, T. and {Ledos}, N. and {Mao}, H. and {Travascio}, A.},
        title = "{The galaxy-halo connection and the dynamical evolution of a giant disc in a massive node of the Cosmic Web at z\raisebox{-0.5ex}\textasciitilde3}",
      journal = {arXiv e-prints},
         year = 2026,
        month = may,
          eid = {arXiv:2605.04144},
        pages = {arXiv:2605.04144},
          doi = {10.48550/arXiv.2605.04144},
archivePrefix = {arXiv},
       eprint = {2605.04144},
 primaryClass = {astro-ph.GA},
       adsurl = {https://ui.adsabs.harvard.edu/abs/2026arXiv260504144Q}
}

@ARTICLE{Robertson23,
       author = {{Robertson}, Brant E. and {Tacchella}, Sandro and {Johnson}, Benjamin D. and {Hausen}, Ryan and {Alabi}, Adebusola B. and {Boyett}, Kristan and {Bunker}, Andrew J. and {Carniani}, Stefano and {Egami}, Eiichi and {Eisenstein}, Daniel J. and {Hainline}, Kevin N. and {Helton}, Jakob M. and {Ji}, Zhiyuan and {Kumari}, Nimisha and {Lyu}, Jianwei and {Maiolino}, Roberto and {Nelson}, Erica J. and {Rieke}, Marcia J. and {Shivaei}, Irene and {Sun}, Fengwu and {{\"U}bler}, Hannah and {Williams}, Christina C. and {Willmer}, Christopher N.~A. and {Witstok}, Joris},
        title = "{Morpheus Reveals Distant Disk Galaxy Morphologies with JWST: The First AI/ML Analysis of JWST Images}",
      journal = {\apjl},
         year = 2023,
        month = jan,
       volume = {942},
       number = {2},
          eid = {L42},
        pages = {L42},
          doi = {10.3847/2041-8213/aca086},
archivePrefix = {arXiv},
       eprint = {2208.11456},
 primaryClass = {astro-ph.GA},
       adsurl = {https://ui.adsabs.harvard.edu/abs/2023ApJ...942L..42R}
}

@ARTICLE{Salo15,
       author = {{Salo}, Heikki and {Laurikainen}, Eija and {Laine}, Jarkko and {Comer{\'o}n}, Sebastien and {Gadotti}, Dimitri A. and {Buta}, Ron and {Sheth}, Kartik and {Zaritsky}, Dennis and {Ho}, Luis and {Knapen}, Johan and et al.},
        title = "{The Spitzer Survey of Stellar Structure in Galaxies (S$^{4}$G): Multi-component Decomposition Strategies and Data Release}",
      journal = {\apjs},
         year = 2015,
        month = jul,
       volume = {219},
       number = {1},
          eid = {4},
        pages = {4},
          doi = {10.1088/0067-0049/219/1/4},
archivePrefix = {arXiv},
       eprint = {1503.06550},
 primaryClass = {astro-ph.GA},
       adsurl = {https://ui.adsabs.harvard.edu/abs/2015ApJS..219....4S}
}

@ARTICLE{Schultheis25,
       author = {{Schultheis}, Mathias and {Sormani}, Mattia C. and {Gadotti}, Dimitri A.},
        title = "{Nuclear stellar discs}",
      journal = {\aapr},
         year = 2025,
        month = nov,
       volume = {33},
       number = {1},
          eid = {7},
        pages = {7},
          doi = {10.1007/s00159-025-00163-6},
archivePrefix = {arXiv},
       eprint = {2509.04562},
 primaryClass = {astro-ph.GA},
       adsurl = {https://ui.adsabs.harvard.edu/abs/2025A&ARv..33....7S}
}

@ARTICLE{Sersic63,
       author = {{S{\'e}rsic}, J.~L.},
        title = "{Influence of the atmospheric and instrumental dispersion on the brightness distribution in a galaxy}",
      journal = {Boletin de la Asociacion Argentina de Astronomia La Plata Argentina},
         year = 1963,
        month = feb,
       volume = {6},
        pages = {41-43},
       adsurl = {https://ui.adsabs.harvard.edu/abs/1963BAAA....6...41S}
}

@software{sklearn25,
  author = {{The scikit-learn developers}},
  title = {scikit-learn},
  year = {2025},
  version = {1.6.1},
  publisher = {Zenodo},
  doi = {10.5281/zenodo.14627164},
  url = {https://doi.org/10.5281/zenodo.14627164},
  note = {Computer software},
}

@ARTICLE{Stacey25,
       author = {{Stacey}, H.~R. and {Kaasinen}, M. and {O'Riordan}, C.~M. and {McKean}, J.~P. and {Powell}, D.~M. and {Rizzo}, F.},
        title = "{A nuclear spiral in a dusty star-forming galaxy at z = 2.78}",
      journal = {\aap},
         year = 2025,
        month = jan,
       volume = {693},
          eid = {L17},
        pages = {L17},
          doi = {10.1051/0004-6361/202452518},
archivePrefix = {arXiv},
       eprint = {2412.03644},
 primaryClass = {astro-ph.GA},
       adsurl = {https://ui.adsabs.harvard.edu/abs/2025A&A...693L..17S}
}

@ARTICLE{Sun24,
       author = {{Sun}, Wen and {Ho}, Luis C. and {Zhuang}, Ming-Yang and {Ma}, Chao and {Chen}, Changhao and {Li}, Ruancun},
        title = "{The Structure and Morphology of Galaxies during the Epoch of Reionization Revealed by JWST}",
      journal = {\apj},
         year = 2024,
        month = jan,
       volume = {960},
       number = {2},
          eid = {104},
        pages = {104},
          doi = {10.3847/1538-4357/acf1f6},
archivePrefix = {arXiv},
       eprint = {2308.09076},
 primaryClass = {astro-ph.GA},
       adsurl = {https://ui.adsabs.harvard.edu/abs/2024ApJ...960..104S}
}

@ARTICLE{Tacchella19,
       author = {{Tacchella}, Sandro and {Diemer}, Benedikt and {Hernquist}, Lars and {Genel}, Shy and {Marinacci}, Federico and {Nelson}, Dylan and {Pillepich}, Annalisa and {Rodriguez-Gomez}, Vicente and {Sales}, Laura V. and {Springel}, Volker and et al.},
        title = "{Morphology and star formation in IllustrisTNG: the build-up of spheroids and discs}",
      journal = {\mnras},
         year = 2019,
        month = aug,
       volume = {487},
       number = {4},
        pages = {5416-5440},
          doi = {10.1093/mnras/stz1657},
archivePrefix = {arXiv},
       eprint = {1904.12860},
 primaryClass = {astro-ph.GA},
       adsurl = {https://ui.adsabs.harvard.edu/abs/2019MNRAS.487.5416T}
}

@ARTICLE{Tacchella16,
       author = {{Tacchella}, Sandro and {Dekel}, Avishai and {Carollo}, C. Marcella and {Ceverino}, Daniel and {DeGraf}, Colin and {Lapiner}, Sharon and {Mandelker}, Nir and {Primack Joel}, R.},
        title = "{The confinement of star-forming galaxies into a main sequence through episodes of gas compaction, depletion and replenishment}",
      journal = {\mnras},
         year = 2016,
        month = apr,
       volume = {457},
       number = {3},
        pages = {2790-2813},
          doi = {10.1093/mnras/stw131},
archivePrefix = {arXiv},
       eprint = {1509.02529},
 primaryClass = {astro-ph.GA},
       adsurl = {https://ui.adsabs.harvard.edu/abs/2016MNRAS.457.2790T}
}

@ARTICLE{Tomassetti16,
       author = {{Tomassetti}, Matteo and {Dekel}, Avishai and {Mandelker}, Nir and {Ceverino}, Daniel and {Lapiner}, Sharon and {Faber}, Sandra and {Kneller}, Omer and {Primack}, Joel and {Sai}, Tanmayi},
        title = "{Evolution of galaxy shapes from prolate to oblate through compaction events}",
      journal = {\mnras},
         year = 2016,
        month = jun,
       volume = {458},
       number = {4},
        pages = {4477-4497},
          doi = {10.1093/mnras/stw606},
archivePrefix = {arXiv},
       eprint = {1512.06268},
 primaryClass = {astro-ph.GA},
       adsurl = {https://ui.adsabs.harvard.edu/abs/2016MNRAS.458.4477T}
}

@article{Umehata24,
    author = "Umehata, Hideki and others",
    title = "{ADF22-WEB: A giant barred spiral starburst galaxy in the z{\,}={\,}3.1 SSA22 protocluster core}",
    eprint = "2410.22155",
    archivePrefix = "arXiv",
    primaryClass = "astro-ph.GA",
    doi = "10.1093/pasj/psaf010",
    journal = "Publ. Astron. Soc. Jap.",
    volume = "77",
    number = "2",
    pages = "432--445",
    year = "2025"
}

@ARTICLE{Vega-Ferrero24,
       author = {{Vega-Ferrero}, Jes{\'u}s and {Huertas-Company}, Marc and {Costantin}, Luca and {P{\'e}rez-Gonz{\'a}lez}, Pablo G. and {Sarmiento}, Regina and {Kartaltepe}, Jeyhan S. and {Pillepich}, Annalisa and {Bagley}, Micaela B. and {Finkelstein}, Steven L. and {McGrath}, Elizabeth J. and {Knapen}, Johan H. and {Arrabal Haro}, Pablo and {Bell}, Eric F. and {Buitrago}, Fernando and {Calabr{\`o}}, Antonello and {Dekel}, Avishai and {Dickinson}, Mark and {Dom{\'\i}nguez S{\'a}nchez}, Helena and {Elbaz}, David and {Ferguson}, Henry C. and {Giavalisco}, Mauro and {Holwerda}, Benne W. and {Kocesvski}, Dale D. and {Koekemoer}, Anton M. and {Pandya}, Viraj and {Papovich}, Casey and {Pirzkal}, Nor and {Primack}, Joel and {Yung}, L.~Y. Aaron},
        title = "{On the Nature of Disks at High Redshift Seen by JWST/CEERS with Contrastive Learning and Cosmological Simulations}",
      journal = {\apj},
         year = 2024,
        month = jan,
       volume = {961},
       number = {1},
          eid = {51},
        pages = {51},
          doi = {10.3847/1538-4357/ad05bb},
archivePrefix = {arXiv},
       eprint = {2302.07277},
 primaryClass = {astro-ph.CO},
       adsurl = {https://ui.adsabs.harvard.edu/abs/2024ApJ...961...51V}
}

@ARTICLE{Wang25,
       author = {{Wang}, Weichen and {Cantalupo}, Sebastiano and {Pensabene}, Antonio and {Galbiati}, Marta and {Travascio}, Andrea and {Steidel}, Charles C. and {Maseda}, Michael V. and {Pezzulli}, Gabriele and {de Beer}, Stephanie and {Fossati}, Matteo and {Fumagalli}, Michele and {Gallego}, Sofia G. and {Lazeyras}, Titouan and {Mackenzie}, Ruari and {Matthee}, Jorryt and {Nanayakkara}, Themiya and {Quadri}, Giada},
        title = "{A giant disk galaxy two billion years after the Big Bang}",
      journal = {Nature Astronomy},
         year = 2025,
        month = may,
       volume = {9},
        pages = {710-719},
          doi = {10.1038/s41550-025-02500-2},
archivePrefix = {arXiv},
       eprint = {2409.17956},
 primaryClass = {astro-ph.GA},
       adsurl = {https://ui.adsabs.harvard.edu/abs/2025NatAs...9..710W}
}

@ARTICLE{Yang26,
       author = {{Yang}, Lilan and {Kartaltepe}, Jeyhan S. and {Drakos}, Nicole E. and {Faisst}, Andreas L. and {Flayhart}, Carter and {Franco}, Maximilien and {Koekemoer}, Anton M. and {Ilbert}, Olivier and {Akins}, Hollis B. and {Casey}, Caitlin M. and {Ding}, Xuheng and {Hadi}, Ali and {Harish}, Santosh and {Laishram}, Ronaldo and {Liu}, Daizhong and {Magdis}, Georgios E. and {Martinez}, III, Felix and {McCracken}, Henry Joy and {Paquereau}, Louise and {Rhodes}, Jason and {Robertson}, Brant E. and {Shuntov}, Marko and {Toni}, Greta},
        title = "{COSMOS-Web: A Multi-wavelength Morphological Catalog of \raisebox{-0.5ex}\textasciitilde780,000 Galaxies}",
      journal = {arXiv e-prints},
         year = 2026,
        month = jun,
          eid = {arXiv:2606.14869},
        pages = {arXiv:2606.14869},
          doi = {10.48550/arXiv.2606.14869},
archivePrefix = {arXiv},
       eprint = {2606.14869},
 primaryClass = {astro-ph.GA},
       adsurl = {https://ui.adsabs.harvard.edu/abs/2026arXiv260614869Y}
}

@ARTICLE{Yu25,
       author = {{Yu}, Si-Yue and {Xu}, Dewang and {Kalita}, Boris S. and {Li}, Sijia and {Silverman}, John D. and {Liang}, Xinyue and {Fang}, Taotao},
        title = "{Color profiles of disk galaxies at z = 1─3 observed with JWST: Implications for outer-disk formation histories}",
      journal = {\aap},
         year = 2025,
        month = jan,
       volume = {693},
          eid = {L9},
        pages = {L9},
          doi = {10.1051/0004-6361/202452752},
archivePrefix = {arXiv},
       eprint = {2412.13064},
 primaryClass = {astro-ph.GA},
       adsurl = {https://ui.adsabs.harvard.edu/abs/2025A&A...693L...9Y}
}
\bibliographystyle{aasjournalv7}



\end{document}